%% file: ESE-JHipster-main.tex
\RequirePackage{fix-cm}
\documentclass[twocolumn]{svjour3}          
\smartqed  
\usepackage[T1]{fontenc} 
\usepackage[utf8]{inputenc}
\usepackage{graphicx}
\usepackage{url}
\usepackage{cite}
\usepackage{paralist}
\usepackage[table]{xcolor}
\rowcolors{1}{}{gray!15}
\usepackage{tabularx}
\usepackage{subfigure}
\usepackage{todonotes}
\usepackage{xspace}
\usepackage{listings}
\usepackage{textcomp}
\usepackage{balance}

\usepackage{tcolorbox}

\newcommand{\etal}{\textit{et al.}\xspace}

\journalname{Empirical Software Engineering}

\begin{document}

\title{Test them all, is it worth it? Assessing configuration sampling on the JHipster Web development stack}


\author{Axel Halin \and
        Alexandre Nuttinck \and
        Mathieu Acher \and 
        Xavier Devroey \and
        Gilles Perrouin \and
        Benoit Baudry
}


\institute{A. Halin, G. Perrouin (FNRS research associate) \at
              PReCISE, NaDI, Faculty of Computer Science\\
              University of Namur, Belgium\\
              \email{gilles.perrouin@unamur.be}           
           \and
           A. Nuttinck \at
              CETIC, Belgium \\
              \email{alexandre.nuttinck@cetic.be}
           \and
           M. Acher \at
              IRISA, University of Rennes I, France \\
              \email{mathieu.acher@irisa.fr}
           \and
           X. Devroey \at
              SERG, Delft University of Technology, The Netherlands\\
              \email{x.d.m.devroey@tudelft.nl}
           \and
           B. Baudry \at
             KTH Royal Institute of Technology, Sweden \\
             \email{baudry@kth.se}
}

\date{Received: date / Accepted: date}

\maketitle

\begin{abstract}
Many approaches for testing configurable software systems start from the same assumption: it is impossible to test all configurations. This motivated the definition of variability-aware abstractions and sampling techniques to cope with large configuration spaces. Yet, there is no theoretical barrier that prevents the exhaustive testing of all configurations by simply enumerating them if the effort required to do so remains acceptable. Not only this: we believe there is a lot to be learned by systematically and exhaustively testing a configurable system. 
In this case study, we report on the first ever endeavour to test all possible configurations of the industry-strength, open source configurable software system: JHipster, a popular code generator for web applications. We built a testing scaffold for the 26,000+ configurations of JHipster using a cluster of 80 machines during 4 nights for a total of 4,376 hours (182 days) CPU time. We find that 35.70\% configurations fail and we identify the feature interactions that cause the errors. We show that sampling strategies (like dissimilarity and 2-wise): (1) are more effective to find faults than the 12 default configurations used in the JHipster continuous integration; (2) can be too costly and exceed the available testing budget. We cross this quantitative analysis with the qualitative assessment of JHipster's lead developers.
\end{abstract}


\input{parts_introduction}

\input{parts_background}

\input{parts_case}

\input{parts_evaluation}

\input{parts_otherstudies}
\input{parts_viewpoint}
\input{parts_threats}

\input{parts_conclusion}

\begin{acknowledgements}
We would like to thank Prof. Arnaud Blouin for his comments and feedback on the paper.
This research was partially funded by the EU Project STAMP ICT-16-10 No.731529 and the Dutch 4TU project ``Big Software on the Run'' as well as by the European Regional Development Fund (ERDF) ``Ideas for the future Internet" (IDEES) project.    
\end{acknowledgements}


\balance

\bibliographystyle{spmpsci}      
\bibliography{references2}   

\end{document}

%% file: parts_introduction.tex
\section{Introduction}

Configurable systems offer numerous options (or features) that promise to fit the needs of different users. New functionalities can be activated or deactivated and some technologies can be replaced by others for addressing a diversity of deployment contexts, usages, etc. The engineering of highly-configurable systems is a standing goal of numerous software projects but it also has a significant cost in terms of development, maintenance, and testing. 
A major challenge for developers of configurable systems is to ensure that all combinations of options (configurations) correctly compile, build, and run. Configurations that fail can hurt potential users, miss opportunities, and degrade the success or reputation of a project. Ensuring quality for all configurations is a difficult task.  For example, Melo \etal compiled 42,000+ random Linux kernels and found that only 226 did not yield any compilation warning \cite{Melo:2016:QAV:2866614.2866615}. Though formal methods and program analysis can identify some classes of defects~\cite{DBLP:journals/csur/ThumAKSS14,Classen2013b} -- leading to variability-aware testing approaches (e.g., \cite{nguyen2014exploring,kim2013splat,kim2011reducing}) -- a common practice is still to execute and \emph{test} a sample of (representative) configurations.  Indeed, enumerating all configurations is perceived as impossible, impractical or both. While this is generally true, we believe there is a lot to be learned by rigorously and exhaustively testing a configurable system. Prior empirical investigations (e.g., \cite{MKRGA:ICSE16,Sanchez2013,Sanchez2017}) suggest that using a sample of configurations is effective to find configuration faults, at low cost. However, evaluations were carried out on a small subset of the total number of configurations or faults, constituting a threat to validity. They typically rely on a corpus of faults that are mined from issue tracking systems.  Knowing all the failures of the whole configurable system provides a unique opportunity to accurately assess the error-detection capabilities of sampling techniques \emph{with a ground truth}. Another limitation of prior works is that the cost of testing configurations can only be estimated. They generally ignore the exact computational cost (e.g., time needed) or how difficult it is to instrument testing for any configuration. 
 
This article aims to grow the body of knowledge (e.g., in the fields of combinatorial testing and software product line engineering~\cite{MKRGA:ICSE16,MWKTS:ASE16,Hervieu2011,Henard2014a,Cohen2008,Sanchez2013}) with a new research approach: the exhaustive testing of \textbf{\it all} configurations. We use JHipster, a popular code generator for web applications, as a case study. Our goals are: 
\begin{inparaenum}[(i)]
\item to investigate the engineering effort and the computational resources needed for deriving and testing all configurations, and
\item to discover how many failures and faults can be found using exhaustive testing in order to provide a ground truth for comparison of diverse testing strategies.
\end{inparaenum}
We describe the efforts required to distribute the testing scaffold for the 26,000+ configurations of JHipster, as well as the interaction bugs that we
discovered. We cross this analysis with the qualitative assessment of JHipster's lead developers. Overall, we collect multiple sources that are of interest for 
\begin{inparaenum}[(i)]
\item researchers interested in building evidence-based theories or tools for testing configurable systems;
\item practitioners in charge of establishing a suitable strategy for testing their systems at each commit or release.
\end{inparaenum}
This article builds on preliminary results \cite{Halin2017} that introduced the JHipster case for research in configurable systems and described early experiments with the testing infrastructure on a very limited number of configurations (300). In addition to providing a quantitative assessment of sampling techniques on \textbf{\textit{all}} the configurations, the present contribution presents numerous qualitative and quantitative insights on building the testing infrastructure itself and compares them with JHipster developers' current practice. In short, we report on the first ever endeavour to test all possible
configurations of the industry-strength open-source configurable software system: JHipster. While there have been efforts in this direction for Linux kernels, their variability space forces to focus on subsets (the selection of 42,000+ kernels corresponds to one month of computation \cite{Melo:2016:QAV:2866614.2866615}) or to investigate bugs qualitatively \cite{Abal:2014,DBLP:journals/tosem/AbalMSBRW18}.  Specifically, the main contributions and findings of this article are:
\begin{enumerate}
\item a cost assessment and qualitative insights of engineering an infrastructure able to automatically test all configurations. This infrastructure is itself a configurable system and requires a substantial, error-prone, and iterative effort (8 man*month);
\item a computational cost assessment of testing all configurations using a cluster of distributed machines. Despite some optimizations, 4,376 hours ($\sim$182 days) CPU time and 5.2 terabytes of available disk space are needed to execute 26,257 configurations; 
\item a quantitative and qualitative analysis of failures and faults. We found that 35.70\% of all configurations fail: they either do not compile, cannot be built or fail to run. Six feature interactions (up to 4-wise) mostly explain this high percentage;
\item an assessment of sampling techniques. Dissimilarity and t-wise sampling techniques are effective to find faults that cause a lot of failures while requiring small samples of configurations. Studying both fault and failure efficiencies provides a more nuanced perspective on the compared techniques; 
\item a retrospective analysis of JHipster practice. The 12 configurations used in the continuous integration for testing JHipster were not able to find the defects. It took several weeks for the community to discover and fix the 6 faults;
\item a discussion on the future of JHipster testing based on collected evidence and feedback from JHipster's lead developers;
\item a feature model for JHipster v3.6.1 and a dataset to perform ground truth comparison of  configuration sampling techniques, both available at \url{https://github.com/xdevroey/jhipster-dataset}.
\end{enumerate} 
 
The remainder of this article is organised as follows: Section \ref{sec:background} provides background information on sampling techniques and motivates the case; Section \ref{sec:case} presents the JHipster case study, the research questions, and methodology applied in this article; Section \ref{sec:testingcosts} presents the human and computational cost of testing all JHipster configurations; Section \ref{sec:rq2} presents the faults and failures found during JHipster testing; Section \ref{subsec:jhipstersampling} makes a ground truth comparison of the sampling strategies; Section \ref{sec:otherstudied} positions our approach with respect to studies comparing sampling strategies on other configurable systems; Section \ref{sec:practitioner} gives the practitioners point of view on JHipster testing by presenting the results of our interview with JHipster developers; Section \ref{sec:threats} discusses the threats to validity; and Section \ref{sec:conclusion} wraps up with conclusions.

%% file: parts_background.tex

\section{Background and Related Work}
\label{sec:background}

\begin{figure*}[t]
  \centering
	\includegraphics[width=\textwidth]{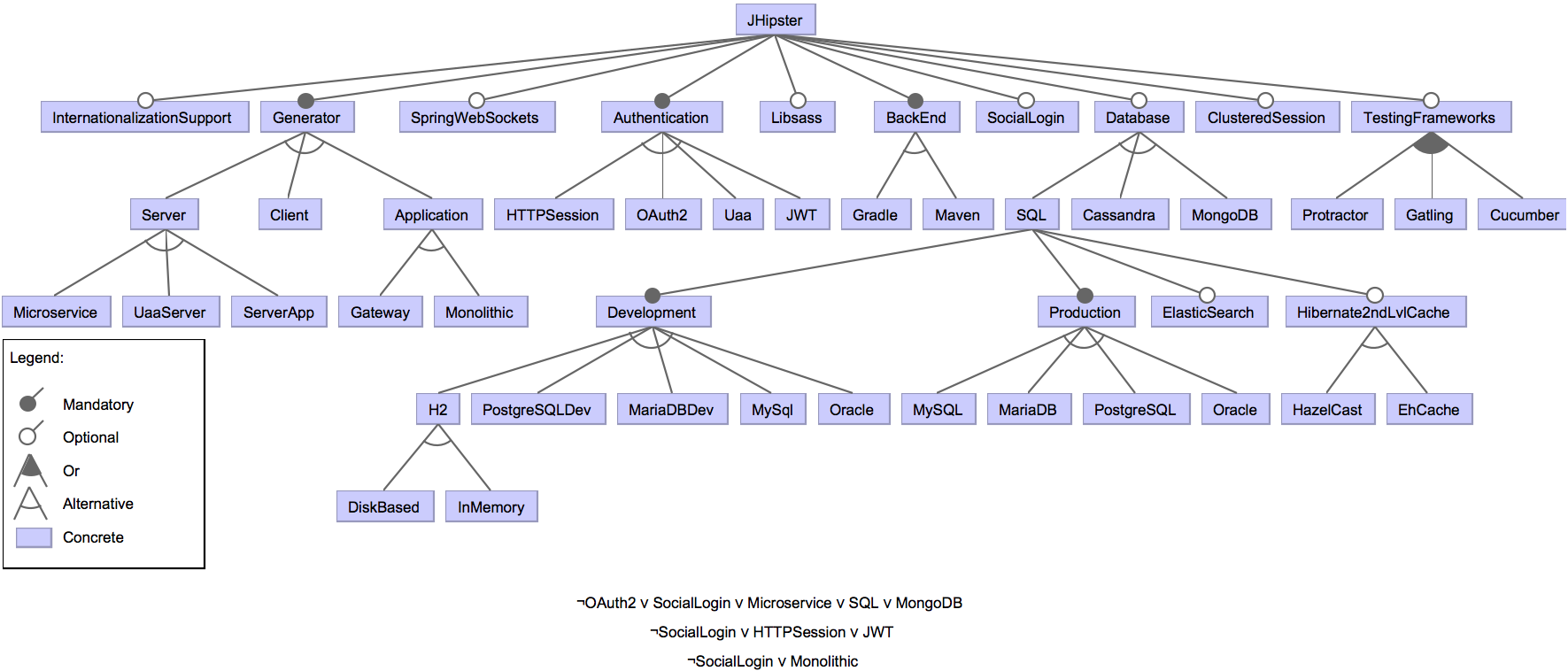}
    \caption{JHipster reverse engineered feature model (only an excerpt of cross-tree constraints is given).}
    \label{fig:fullfm}
\end{figure*}

Configurable systems have long been studied by the Software Product Line (SPL) engineering community \cite{Pohl2005,Apel2013}. They use a tree-like structure, called \emph{feature model} \cite{Kang1990}, to represent the set of valid combinations of options: i.e., the variants (also called products). Each option (or \emph{features}\footnote{In the remaining of this paper, we consider \emph{features} as units of variability: i.e., \emph{options}.}) maybe decomposed into sub-features and additional constraints may be specified amongst the different features. 

For instance, Figure \ref{fig:fullfm} presents the full feature model of JHipster. Each JHipster variant has a \textsf{Generator} option that may be either a \textsf{Server}, a \textsf{Client}, or an \textsf{Application}; may also have a \textsf{Database} that is \textsf{SQL} or \textsf{Cassandra} or \textsf{MongoDB}; etc. Additional constraints specify for instance that \textsf{SocialLogin} may only be selected for \textsf{Monolithic} applications.

\subsection{Reverse Engineering Variability Models}

The first step required to reason on an existing configurable system is to identify its variability. There are some approaches in the literature that attempt to extract variability and synthesize a feature model. For example, She \etal devised a technique to transform the description language of the Linux kernel into a representative feature model \cite{She:2011:REF:1985793.1985856}. The inference of parent-child relationships amongst features proved to be problematic as the well as the mapping of multi-valued options to boolean features.  As a result, feature models extracted with such a technique have to be further validated and corrected \cite{Henard:2013:TAT:2486788.2486975}. Abbasi \etal \cite{DBLP:conf/csmr/AbbasiAHC14} designed an extraction approach that first look for variability patterns in web configurator tools and complete extracted information using a web crawler. In this case, the feature model is not synthesised. Indeed, static analysis has been largely used to reason about configuration options at the code level (e.g., \cite{Rabkin:2011:SEP:1985793.1985812,Nadi2015}). Such techniques often lie at the core of variability-aware testing approaches discussed below. As we will detail in our study, the configurator implementation as well as variation points of JHipster are scattered in different kinds of artefacts, challenging the use of static and dynamic analyses. As a result, we rather used a manual approach to extract a variability model. Though automated variability extraction can be interesting to study JHipster evolution over the long term, we leave it out of the scope of the present study.

\subsection{Testing a Configurable System}

Over the years, various approaches have been developed to test configurable systems~\cite{DaMotaSilveiraNeto2011,Engstrom2011,Machado2014}. They can be classified into two strategies: \emph{configurations sampling} and \emph{variability-aware testing}. Configuration sampling approaches sample a representative subset of all the valid configurations of the system and test them individually. Variability-aware testing approaches instrument the testing environment to take variability information and reduce the test execution effort.

\subsubsection{Variability-aware testing}

To avoid re-execution of variants that have exactly the same execution paths for a test case, Kim \etal and Shi \etal use static and dynamic execution analysis to collect variability information from the different code artefacts and remove relevant configurations accordingly ~\cite{kim2013splat,Shi2012}. 

Variability-aware execution approaches \cite{kim2011reducing,nguyen2014exploring,Austin2012} instrument an interpreter of the underlying programming language to execute the tests only once on all the variants of a configurable system. For instance, Nguyen \etal implemented Varex, a variability-aware PHP interpreter, to test WordPress by running code common to several variants only once \cite{nguyen2014exploring}. Alternatively, instead of executing the code, Reisner \etal use a symbolic execution framework to evaluate how the configuration options impact the coverage of the system for a given test suite \cite{Reisner2010}. Static analysis and notably type-checking has been used to look for bugs in configurable software~\cite{kastner2008type,Kenner2010}. A key point of type-checking approaches is that they have been scaled to very large code bases such as the Linux kernel.    

Although we believe that JHipster is an interesting candidate case study for those approaches, with the extra difficulty that variability information is scattered amongst different artefacts written in different languages (as we will see in Section \ref{sec:REV}), they require a (sometimes heavy) instrumentation of the testing environment. Therefore, we leave variability-aware testing approaches outside the scope of this case study and focus instead on configuration sampling techniques that can fit into the existing continuous integration environment of JHipster developers (see Section \ref{subsec:jhipsterTesting}). 


\subsubsection{Configurations sampling}
\label{subsec:backgroundsampling}

\paragraph{Random sampling.}
This strategy is straightforward: select a random subset of the valid configurations. Arcuri \etal~\cite{Arcuri2012} demonstrate that, in the absence of constraints between the options, this sampling strategy may outperform other sampling strategies. In our evaluation, random sampling serves as basis for comparison with other strategies.


\paragraph{T-wise sampling.}
%
T-wise sampling comes from Combinatorial Interaction Testing (CIT), which relies on the hypothesis that most faults are caused by undesired interactions of a small number of features \cite{Kuhn2004}. This technique has been adapted to variability-intensive systems for more than 10 years \cite{Cohen2008,Lopez-Herrejon2015a}.  
A t-wise algorithm samples a set of configurations such that all possible $t$-uples of options are represented at least once (it is generally not possible to have each $t$-uples represented exactly once due to constraints between options). Parameter $t$ is called \textit{interaction strength}. The most common t-wise sampling is pairwise (2-wise)~\cite{yilmaz2006covering,Cohen2008,Perrouin2011,Johansen2016,Hervieu2011}. In our evaluation, we rely on SPLCAT~\cite{Johansen2012}, an efficient t-wise sampling tool for configurable systems based on a greedy algorithm.

\paragraph{Dissimilarity sampling.}
Despite advances being made, introducing constraints during t-wise sampling yields scalability issues for large feature models and higher interaction strengths~\cite{MKRGA:ICSE16}. To overcome those limitations, Henard \etal developed a dissimilarity-driven sampling ~\cite{Henard2014a}. This technique approximates t-wise coverage by generating dissimilar configurations (in terms of shared options amongst these configurations). From a set of random configurations of a specified cardinality, a (1+1) evolutionary algorithm evolves this set such that the distances amongst configurations are maximal, by replacing a configuration at each iteration, within a certain amount of time. In our evaluation, we rely on Henard \etal's implementation: PLEDGE~\cite{Henard2013PLEDGE}. The relevance of dissimilarity-driven sampling for software product lines has been empirically demonstrated for large feature models and higher strengths~\cite{Henard2014a}. This relevance was also independently confirmed for smaller SPLs~\cite{Al-Hajjaji2016}.

\paragraph{Incremental Sampling}
Incremental sampling consists of focusing on one configuration and progressively adding new ones that are related to focus on specific parts of the configuration space \cite{uzuncaova2010incremental,oster2010automated,delta-MBT}. For example, Lochau \etal \cite{delta-MBT} proposed a model-based approach that  shifts from one product to another  by applying ``deltas'' to statemachine models. These deltas enable automatic reuse/adaptation of test model and derivation of retest obligations. Oster \etal extend combinatorial interaction testing with the possibility to specify a predefined set of products in the configuration suite to be tested \cite{oster2010automated}. Incremental techniques naturally raise the issue of which configuration to start from. Our goal was to compare techniques that explore the configuration space in the large and therefore we did not include incremental techniques in our experiments.  

\paragraph{One-disabled sampling.}
The core idea of one-di\-sa\-bled sampling is to extract configurations in which all options are activated but one \cite{Abal:2014,MKRGA:ICSE16}. For instance, in the feature diagram of Figure \ref{fig:fullfm}, we will have a configuration where the \textsf{SocialLogin} option is deactivated and all the other options (that are not mandatory) are activated. 

This criterion allows various strategies regarding its implementation: in our example, one may select a configuration with a \textsf{Server} xor \textsf{Client} xor \textsf{Application} option active. All those three configurations fit for the one-disabled definition. In their implementation, Medeiros \etal \cite{MKRGA:ICSE16} consider the first valid configuration returned by the solver. 
Since SAT solvers rely on internal orders to process solutions (see \cite{Henard2014a}) the first valid solution will always be the same. The good point is that it makes the algorithm deterministic. However, it implicitly links the bug-finding ability of the algorithm with the solver's internal order and to the best of our knowledge, there is no reason why it should be linked. 

In our evaluation (see Section \ref{subsec:twisesampling}), for each disabled option, we choose to apply a random selection of the configuration to consider. Additionally, we also extend this sampling criteria to all valid configurations where one feature is disabled and the others are enabled (called \emph{all-one-disabled} in our results): in our example, for the \textsf{SocialLogin} option deactivated, we will have one configuration with \textsf{Server} option activated, one configuration with \textsf{Client} option activated, and one configuration with \textsf{Application} option activated.

\paragraph{One-enabled sampling.}
This sampling mirrors one-di\-sa\-bled and consists of enabling each option one at a time \cite{Abal:2014,MKRGA:ICSE16}. For instance, a configuration where the \textsf{SocialLogin} option is selected and all the other options are deselected. As for one-disabled, for each selected option, we apply a random selection of the configuration to consider in our evaluation; and the criteria are extended to \emph{all-one-enabled}, with all the valid configurations for each selected option.

\paragraph{Most-enabled-disabled sampling.}
This method only samples two configurations: one where as many options as possible are selected and one where as many options as possible are deselected \cite{Abal:2014,MKRGA:ICSE16}. If more than one valid configuration is possible for most-enabled (respectively most-disabled) options, we randomly select one most-enabled (respectively most-disabled) configuration. The criteria are extended to \emph{all-most-enabled-disabled}, with all the valid configurations with most-enabled (respectively most-disabled) options.

\paragraph{Other samplings.}
Over the years, many other sampling techniques have been developed. Some of them use other artefacts in combination with the feature model to perform the selection. 
Johansen \etal~\cite{DBLP:conf/models/JohansenHFES12} extended SPL\-CAT by adding weights on sub-product lines. 
Lochau \etal combine coverage of the feature model with test model coverage, such as control and data flow coverage~\cite{Lochau2011}.
Devroey \etal switched the focus from variability to behaviour \cite{Devroey2014e,Devroey2016} and usage of the system \cite{Devroey2015a} by considering a featured transition system for behaviour and configurations sampling.
%
%
In this case study, we only consider the feature model as input for our samplings and focus on random, t-wise, dissimilarity, one-enabled, one-disabled, and most-enabled-disabled techniques. 

\subsection{Comparison of Sampling Approaches}


Perrouin \etal \cite{Perrouin2011} compared two exact approaches on five feature models of the SPLOT repository w.r.t to performance of t-wise generation and configuration diversity. 
Hervieu \etal \cite{Hervieu2011} also used models from the SPLOT repository to produce a small number of configurations. 
Johansen \etal's \cite{DBLP:conf/models/JohansenHFES12} extension of SPLCAT has been applied to the Eclipse IDE and to TOMRA, an industrial product line.
Empirical investigations were pursued on larger models (1,000 features and above) notably on OS kernels (e.g., \cite{Henard2014a,Johansen2012}) demonstrating the relevance of metaheuristics for large sampling tasks \cite{Henard2015,Ochoa:2017:SSP:3023956.3023959}.  
However, these comparisons were performed at the model level using artificial faults. 

Several authors considered sampling on actual systems. Steffens \etal \cite{Oster2011} applied the Moso-Polite pairwise tool on an electronic module allowing 432 configurations to derive metrics regarding the test reduction effort. Additionally, they also exhibited a few cases where a higher interaction strength was required (3-wise). 

Finally, in Section \ref{sec:otherstudied}, we present an in-depth discussion of related case studies with sampling techniques comparison.


\subsection{Motivation of this Study}

Despite the number of empirical investigations (e.g., \cite{ganesan2007comparing,qu2008configuration}) and surveys (e.g., \cite{Engstrom2011,DBLP:journals/csur/ThumAKSS14,DaMotaSilveiraNeto2011}) to compare such approaches, many focused on subsets to make the analyses tractable. 
Being able to execute all configurations led us to consider actual failures and collect a ground truth. It helps to gather insights for better understanding the interactions in large configuration spaces~\cite{MWKTS:ASE16,yilmaz2006covering}.
And provide a complete, open, and reusable dataset to the configurable system testing community to evaluate and compare new approaches.

%% file: parts_case.tex
\section{Case Study}
\label{sec:case}

\emph{JHipster} is an open-source, industrially used generator for developing Web applications~\cite{jhipster:Website}. Started in 2013, the JHipster project has been increasingly popular (6000+ stars on GitHub) with a strong community of users and around 300 contributors in February 2017. 

From a user-specified configuration, JHipster generates a complete technological stack constituted of Java and Spring Boot code (on the server side) and Angular and Bootstrap (on the front-end side). The generator supports several technologies ranging from the database used (e.g., \textit{MySQL} or \textit{MongoDB}), the authentication mechanism (e.g., \textit{HTTP Session} or \textit{Oauth2}), the support for social log-in (via existing social networks accounts), to the use of microservices.
Technically, JHipster uses \textit{npm} and \textit{Bower} to manage dependencies and Yeoman\footnote{\url{http://yeoman.io/}} (aka \textit{yo}) tool to scaffold the application \cite{raible:jhipsterBook}. JHipster relies on conditional compilation with EJS\footnote{\url{http://www.embeddedjs.com/}} as a variability realisation mechanism.
Listing~\ref{lst:javaTemplate} presents an excerpt of class \textit{DatabaseConfiguration.java}. The options \textsf{sql}, \textsf{mongodb}, \textsf{h2Disk}, \textsf{h2Memory} operate over Java annotations, fields, methods, etc. For instance, on line \ref{line:javaTEmplate:mongodb}, the inclusion of \textsf{mongodb} in a configuration means that \textsf{DatabaseConfiguration} will inherit from \textsf{AbstractMongoConfiguration}.

\begin{lstlisting}[float=t,frame={single}, caption={Variability in \_DatabaseConfiguration.java},label={lst:javaTemplate},basicstyle=\scriptsize\ttfamily,breaklines=true,numbers=left]
(...)
@Configuration<% if (databaseType == 'sql') { %>
@EnableJpaRepositories("<%=packageName%>.repository")
@EnableJpaAuditing(...)
@EnableTransactionManagement<% } %>
(...)
public class DatabaseConfiguration 
<% if (databaseType == 'mongodb') { %> (*@\label{line:javaTEmplate:mongodb}@*)
	extends AbstractMongoConfiguration 
<% } %>{

 <%_ if (devDatabaseType == 'h2Disk' || devDatabaseType == 'h2Memory') { _%>
   /**
   * Open the TCP port for the H2 database.
   * @return the H2 database TCP server
   * @throws SQLException if the server failed to start
   */
   @Bean(initMethod = "start", destroyMethod = "stop")
   @Profile(Constants.SPRING_PROFILE_DEVELOPMENT)
   public Server h2TCPServer() throws SQLException {
      return Server.createTcpServer(...);
   }
  <%_ } _%>
(...)
\end{lstlisting}

JHipster is a \emph{complex} configurable system with the following characteristics:
\begin{inparaenum}[(i)]
\item a variety of languages (Ja\-va\-Script, CSS, SQL, etc.) and advanced technologies (Maven, Docker, etc.) are combined to generate variants;
\item there are 48 configuration options and a configurator guides user throughout different questions. Not all combinations of options are possible and there are 15 constraints between options;
\item variability is scattered among numerous kinds of artefacts (\textit{pom.xml}, Java classes, Docker files, etc.) and several options typically contribute to the activation or deactivation of portions of code, which is commonly observed in configurable software \cite{Jin2014}. 
\end{inparaenum}

This complexity challenges core developers and contributors of JHipster. Unsurprisingly, numerous configuration faults have been reported on mailing lists and eventually fixed with commits.\footnote{e.g., \url{https://tinyurl.com/bugjhipster15}}  Though formal methods and variability-aware program analysis can identify some defects~\cite{DBLP:journals/csur/ThumAKSS14,Classen2013b,nguyen2014exploring}, a significant effort would be needed to handle them in this technologically diverse stack. Thus, the current practice is rather to execute and test some configurations and JHipster offers opportunities to assess the cost and effectiveness of sampling strategies~\cite{MKRGA:ICSE16,MWKTS:ASE16,Hervieu2011,Henard2014a,Cohen2008,Sanchez2013}. Due to the reasonable number of options and the presence of 15 constraints, we (as researchers) also have a unique opportunity to gather a ground truth through the testing of \emph{all} configurations.

\subsection{Research Questions}

Our research questions are formulated around three axes: the first one addresses the feasibility of testing all JHipster configurations; the second question addresses the bug-discovery power of state-of-the-art configuration samplings; and the last one addresses confronts our results with the JHipster developers point of view.

\subsubsection{\textbf{(RQ1)} What is the feasibility of testing all JHipster configurations?}

This research question explores the cost of an exhaustive and automated testing strategy. It is further decomposed into two questions:
\begin{itemize}
\item[\textbf{(RQ1.1)}] What is the cost of engineering an infrastructure capable of automatically deriving and testing all configurations?
\end{itemize}
To answer this first question, we reverse engineered a feature model of JHipster based on various code artefacts (described in Section \ref{sec:REV}), and devise an analysis workflow to automatically derive, build, and test JHipster configurations (described in Section \ref{subsec:workflow}). This workflow has been used to answer our second research question:
\begin{itemize}
\item[\textbf{(RQ1.2)}] What are the computational resources needed to test all configurations?
\end{itemize}
To keep a manageable execution time, the workflow has been executed on the INRIA Grid'5000, a large-scale testbed offering a large amount of computational resources~\cite{Balouek2012}. 

Section \ref{sec:computationalcost} describes our engineering efforts in building a fully automated testing infrastructure for all JHipster variants. 
We also evaluate the computational cost of such an exhaustive testing; describe the necessary resources (man-power, time, machines); and report on encountered difficulties as well as lessons learned.

\subsubsection{\textbf{(RQ2)} To what extent can sampling help to discover defects in JHipster?}

We use the term \emph{defect} to refer to either a fault or a failure. A \emph{failure} is an ``undesired effect observed in the system's delivered service''~\cite{Mathur2008,IEEEComputerSociety:2014:GSE:2616205} (e.g., the JHipster configuration fails to compile). We then consider that a \emph{fault} is a cause of failures. As we found in our experiments (see Section~\ref{sec:rq2}), a single fault can explain many configuration failures since the same feature interactions cause the failure.

To compare different sampling approaches, the first step is to characterise failures and faults that can be found in JHipster:
\begin{itemize}
\item[\textbf{(RQ2.1)}] How many and what kinds of failures/faults can be found in all configurations?
\end{itemize}
Based on the outputs of our analysis workflow, we identify the faults causing one or more failures using statistical analysis (see Section \ref{subsec:statisticalbugs}) and confirm those faults using qualitative analysis, based on issue reports of the JHipster GitHub project (see Section \ref{subsec:qualitativebugs}).

By collecting a \emph{ground truth} (or reference) of defects, we can measure the effectiveness of sampling techniques. For example, is a random selection of 50 (says) configurations as effective to find failures/faults than an exhaustive testing? We can address this research question:
\begin{itemize}
\item[\textbf{(RQ2.2)}] How effective are sampling techniques comparatively? 
\end{itemize}
We consider the sampling techniques presented in Section~\ref{subsec:backgroundsampling}; all techniques use the feature model as primary artefact (see Section \ref{subsec:jhipstersampling}) to perform the sampling. For each sampling technique, we measure the failures and the associated faults that the sampled configurations detect. Besides a comparison between automated sampling techniques, we also compare the manual sampling strategy of the JHipster project.

Since our comparison is performed using specific results of JHipster's executions and cannot be generalized as such, we confront our findings to other case studies found in the literature. In short:
\begin{itemize}
\item[\textbf{(RQ2.3)}] How do our sampling techniques effectiveness findings compare to other case studies and works? 
\end{itemize}
To answer this question, we perform a literature review on empirical evaluation of sampling techniques (see Section \ref{sec:otherstudied}).

\subsubsection{\textbf{(RQ3)} How can sampling help JHipster developers?}

Finally, we can put in perspective the typical \emph{trade-off} between the ability to find configuration defects and the cost of testing.
\begin{itemize}
\item[\textbf{(RQ3.1)}] What is the most cost-effective sampling strategy for JHipster?
\end{itemize}
And confront our findings to the current development practices of the JHipster developers:
\begin{itemize}
\item[\textbf{(RQ3.2)}] What are the recommendations for the JHipster project? 
\end{itemize}
To answer this question, we performed a semi-structured interview of the lead developer of the project and exchanged e-mails with other core developers to gain insights on the JHipster development process and collect their reactions to our recommendations, based on an early draft of this paper (see Section \ref{sec:practitioner}).

\subsection{Methodology}

We address these questions through quantitative and qualitative research. We initiated the work in September 2016 and selected JHipster 3.6.1\footnote{\url{https://github.com/jhipster/generator-jhipster/releases/tag/v3.6.1}} (release date: mid-August 2016). The 3.6.1 corrects a few bugs from 3.6.0; the choice of a ``minor'' release avoids finding bugs caused by an early and unstable release.   

The two first authors worked full-time for four months to develop the infrastructure capable of testing all configurations of JHipster. They were graduate students, with strong skills in programming and computer science. Prior to the project's start, they have studied feature models and JHipster. We used GitHub to track the evolution of the testing infrastructure. We also performed numerous physical or virtual meetings (with Slack). Four other people have supervised the effort and provided guidance based on their expertise in software testing and software product line engineering. Through frequent exchanges, we gather several qualitative insights throughout the development.

Besides, we decided not to report faults whenever we found them. Indeed, we wanted to observe whether and how fast the JHipster community would discover and correct these faults.  We monitored JHipster mailing lists to validate our testing infrastructure and characterize the configuration failures in a qualitative way. We have only considered GitHub issues since most of the JHipster activity is there. Additionally, we used statistical tools to quantify the number of defects, as well as to assess sampling techniques. Finally, we crossed our results with insights from three JHipster's lead developers.

%% file: parts_evaluation.tex
\section{All Configurations Testing Costs (RQ1)}
\label{sec:testingcosts}


\subsection{Reverse Engineering Variability}
\label{sec:REV}

The first step towards a complete and thorough testing of JHipster variants is the modelling of its configuration space. JHipster comes with a command-line configurator. However, we quickly noticed that a brute force tries of every possible combinations has scalability issues. Some answers activate or deactivate some questions and options. As a result, we rather considered the source code from GitHub for identifying options and constraints. Though options are scattered amongst artefacts, there is a central place that manages the configurator and then calls different sub-generators to derive a variant.

We essentially consider \textit{prompts.js}, which specifies questions prompted to the user during the configuration phase, possible answers (a.k.a. options), as well as constraints between the different options. Listing~\ref{lst:promptJS} gives an excerpt for the choice of a \textsf{databaseType}. Users can select no database, \textsf{sql}, \textsf{mongodb}, or \textsf{cassandra} options. There is a pre-condition stating that the prompt is presented only if the \textsf{microservice} option has been previously selected (in a previous question related to \textsf{applicationType}). In general, there are several conditions used for basically encoding constraints between options. 

\begin{lstlisting}[float=t,frame={single}, caption={Configurator: server/prompt.js (excerpt)},label={lst:promptJS},basicstyle=\scriptsize\ttfamily,breaklines=true,numbers=left]
(...)
when: function (response) {
    return applicationType === 'microservice';
   },
   type: 'list',
   name: 'databaseType',
   message: function (response) {
       return getNumberedQuestion('Which *type* of database would you like to use?', applicationType === 'microservice');},
   choices: [
       {value: 'no', name: 'No database'},
       {value: 'sql', name: 'SQL (H2, MySQL, MariaDB, PostgreSQL, Oracle)'},
       {value: 'mongodb', name: 'MongoDB'},
       {value: 'cassandra',name: 'Cassandra'}
   ],
   default: 1
(...)
\end{lstlisting}

We modelled JHispter's variability using a feature model (e.g., \cite{Kang1990}) to benefit from state-of-the-art reasoning techniques developed in software product line engineering~\cite{benavides2010,Classen2011,Apel2013,DBLP:journals/csur/ThumAKSS14,FAMILIAR}. Though there is a gap with the configurator specification (see Listing~\ref{lst:promptJS}), we can encode its \emph{configuration semantics} and hierarchically organize options with a feature model. We decided to interpret the meaning of the configurator as follows:
\begin{enumerate}
\item each multiple-choice question is an (abstract) feature. In case of ``yes'' or ``no'' answer, questions are encoded as optional features (e.g., \textsf{databaseType} is optional in Listing~\ref{lst:promptJS});
\item each answer is a concrete feature (e.g., \textsf{sql}, \textsf{mongodb}, or \textsf{cassandra} in Listing~\ref{lst:promptJS}). All answers to questions are exclusive and translated as alternative groups in the feature modelling jargon. A notable exception is the selection of testing frameworks in which several answers can be both selected; we translated them as an Or-group;
\item pre-conditions of questions are translated as constraints between features.  
\end{enumerate}

Based on an in-depth analysis of the source code and attempts with the configurator, we have manually reverse-engineered an initial feature model presented in Figure~\ref{fig:fullfm}: 48 identified features and 15 constraints (we only present four of them in Figure~\ref{fig:fullfm} for the sake of clarity). The total number of valid configurations is 162,508. 

\begin{figure*}[t]
  \centering
	\includegraphics[width=\textwidth]{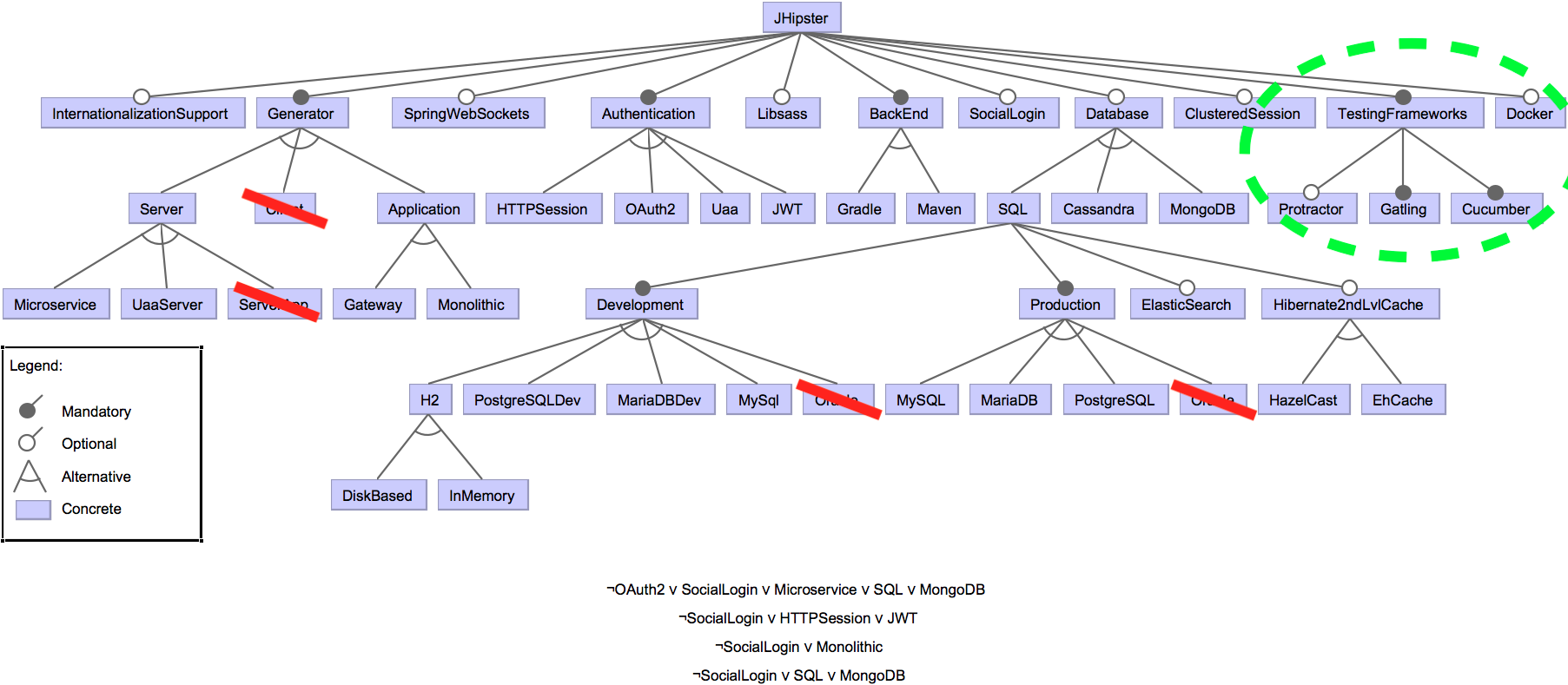}
    \caption{JHipster specialised feature model used to generate JHipster variants  (only an excerpt of cross-tree constraints is given).}
    \label{fig:jhipsterfm}
\end{figure*}

Our goal was to derive and generate all JHipster variants corresponding to feature model configurations. However, we decided to adapt the initial model as follows:
\begin{enumerate}
\item we added Docker as a new optional feature (\textsf{Docker}) to denote the fact that the deployment may be performed using Docker or using Maven or Gradle. Docker has been introduced in JHipster 3.0.0 and is present by default in all generated variants (and therefore does not appear in the feature model of Figure \ref{fig:fullfm}). However, when running JHipster, the user may choose to use it or not, hence the definition of \textsf{Docker} as optional for our analysis workflow: when the option is selected, the analysis workflow performs the deployment using Docker;
\item we excluded client/server standalones since there is a limited interest for users to consider the server (respectively client) without a client (respectively server): stack and failures most likely occur when both sides are inter-related;
\item we included the three testing frameworks in all variants. The three frameworks do not augment the functionality of JHipster and are typically here to improve the testing process, allowing us to gather as much information as possible about the variants;
\item we excluded Oracle-based variants. Oracle is a proprietary technology with technical specificities that are quite hard to fully automate (see Section \ref{subsec:workflow}).
\end{enumerate}

Strictly speaking, we test \emph{all} configurations of a \emph{specialized} JHipster, presented in Figure~\ref{fig:jhipsterfm}. This specialization can be thought of a test model, which focusses on the most relevant open source variants. Overall, we consider that our specialization of the feature model is conservative and still substantial. In the rest of this article, we are considering the original feature model of Figure~\ref{fig:fullfm} augmented with specialized constraints that negate features \textsf{Oracle12c}, \textsf{Oracle}, \textsf{ServerApp}, and \textsf{Client} (in red in Figure~\ref{fig:jhipsterfm}) and that add an optional \textsf{Docker} feature and make \textsf{Gatling} and \textsf{Cucumber} features mandatory (in green in Figure~\ref{fig:jhipsterfm}). 
This specialization leads to a total of \textbf{26,256} variants.

\subsection{Fully Automated Derivation and Testing}
\label{subsec:workflow}

 \begin{figure*}[t]
  \centering
    \includegraphics[width=\textwidth]{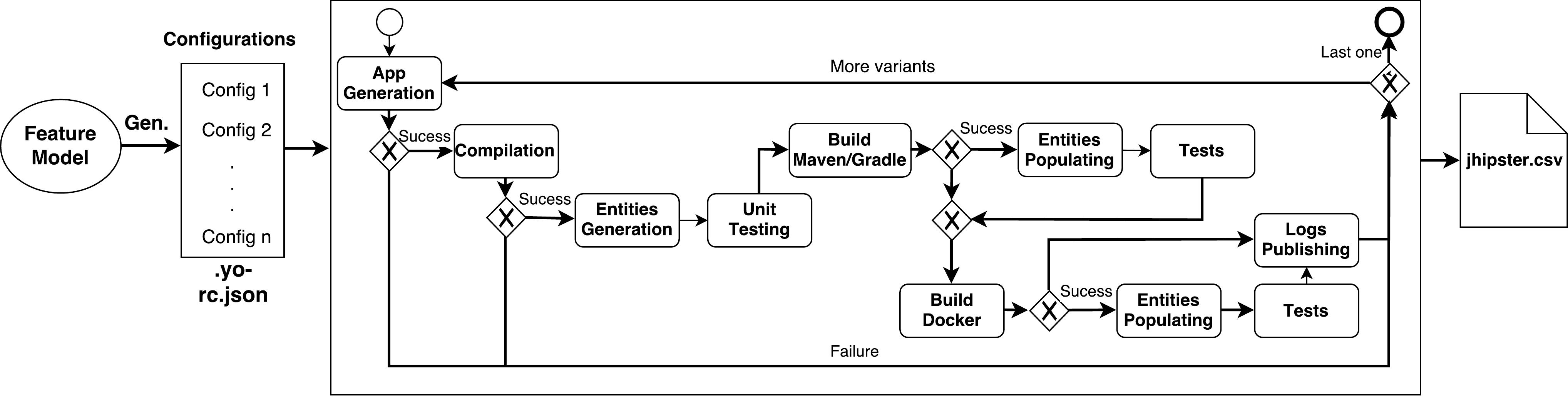}
	\caption{Testing workflow of JHipster configurations.}
    \label{fig:oracleWorkflow}
\end{figure*} 

From the feature model, we enumerated all valid configurations using solvers and FAMILIAR~\cite{FAMILIAR}. We developed a comprehensive workflow for testing each configuration. Figure~\ref{fig:oracleWorkflow} summarises the main steps (compilation, builds and tests). The first step is to synthesize a \textit{.yo-rc.json} file from a feature model configuration. It allows us to skip the command-line questions-and-answers-based configurator; the command \textit{yo jhipster} can directly use such a JSON file for launching the compilation of a variant. A monitoring of the whole testing process is performed to detect and log \emph{failures} that can occur at several steps of the workflow. We faced several difficulties for instrumenting the workflow. 

\subsubsection{Engineering a configurable system for testing configurations} 

The execution of a unique and generic command for testing JHipster variants was not directly possible. For instance, the build of a JHipster application relies either on \textsf{Maven} or \textsf{Gradle}, two alternative features of our variability model. We developed variability-aware scripts to execute commands specific to a JHipster configuration. Command scripts include: starting database services, running database scripts (creation of tables, keyspaces, generation of entities, etc.), launching test commands, starting/stopping Docker, etc. As a concrete example, the inclusion of features \textsf{h2} and \textsf{Maven} lead to the execution of the command: \emph{``mvnw -Pdev"}; the choice of \textsf{Gradle} (instead of \textsf{Maven}) and \textsf{mysql} (instead of \textsf{h2}) in \textsf{production} mode would lead to the execution of another command: \emph{``gradlew -Pprod"}. In total, 15 features of the original feature model influence (individually or through interactions with others) the way the testing workflow is executed. The first lessons learned are that 
\begin{inparaenum}[(i)]
\item a non-trivial engineering effort is needed to build a \emph{configuration-aware testing workflow} -- testing a configurable system like JHipster requires to develop another configurable system; 
\item the development was iterative and mainly consisted in \emph{automating all tasks} originally considered as manual (e.g., starting database services).
\end{inparaenum}

\subsubsection{Implementing testing procedures} 

After a successful build, we can execute and test a JHipster variant. A first challenge is to create the generic conditions (i.e., input data) under which all variants will be executed and tested. Technically, we need to populate Web applications with \emph{entities} (i.e., structured data like tables in an SQL database or documents in MongoDB for instance) to test both the server-side (in charge of accessing and storing data) and the client-side (in charge of presenting data). JHipster entities are created using a domain-specific language called JDL, close to UML class diagram formalism. We decided to reuse the entity model template given by the JHipster team\footnote{\url{https://jhipster.github.io/jdl-studio/}}. We created 3 entity models for \textsf{MongoDB}, \textsf{Cassandra}, and ``others'' because some database technologies vary in terms of JDL expressiveness they can support (e.g., you cannot have relationships between entities with a \textsf{MongoDB} database). 

\begin{lstlisting}[float=t,frame={single}, caption={JHipster generated JUnit test in \texttt{AccountResourceIntTest.java}},label={lst:junitexample}, basicstyle=\scriptsize\ttfamily, breaklines=true]
@Test
public void testAuthenticatedUser() (...) {
    restUserMockMvc.perform(get("/api/authenticate")
        .with(request -> {
            request.setRemoteUser("test");
            return request;})
        .accept(MediaType.APPLICATION_JSON))
        .andExpect(status().isOk())
        .andExpect(content().string("test"));
}
\end{lstlisting}

\begin{lstlisting}[float=t,frame={single}, caption={JHipster generated Karma.js test in \texttt{user.service.spec.ts}},label={lst:karmajsexample}, basicstyle=\scriptsize\ttfamily, breaklines=true]
describe('Component Tests', () => {
  describe('LoginComponent', () => {
    (...)
    it('should redirect user when register', () => {
      // WHEN
      comp.register();
      // THEN
      expect(mockActiveModal.dismissSpy).toHaveBeenCalledWith('to state register');
      expect(mockRouter.navigateSpy).toHaveBeenCalledWith(['/register']);
    });
    (...)
  });
});
\end{lstlisting}

After entities creation with JDL (\textit{Entities Generation} in Figure~\ref{fig:oracleWorkflow}), we run several tests: integration tests written in Java using the \textit{Spring Test Context} framework (see Listing \ref{lst:junitexample} for instance), user interface tests written in JavaScript using the \textit{Karma.js} framework (see Listing \ref{lst:karmajsexample} for instance), etc., and create an executable JHipster variant (\textit{Build Maven/Gradle} in Figure~\ref{fig:oracleWorkflow}). The tests run at this step are automatically generated and include defaults tests common to all JHipster variants and additional tests generated by the JDL entities creation. On average, the Java line coverage is 44.53\% and the JavaScript line coverage is 32.19\%.

We instantiate the generated entities (\textit{Entities Populating} in Figure~\ref{fig:oracleWorkflow}) using the Web user interface through Selenium scripts. We integrate the following testing frameworks to compute additional metrics (\textit{Tests} in Figure~\ref{fig:oracleWorkflow}):  Cucumber, Gatling and Protractor. We also implement generic oracles that analyse and extract log error messages. And finally, repeated two last steps using Docker (\textit{Build Docker}, \textit{Entities Populating}, and \textit{Tests} in Figure~\ref{fig:oracleWorkflow}) before saving the generated log files.
 
Finding commonalities among the testing procedures participates to the engineering of a configuration-aware testing infrastructure. The major difficulty was to develop input data (entities) and test cases (e.g., Selenium scripts) that are \emph{generic} and can be applied to all JHipster variants. 
  
\subsubsection{Building an all-inclusive testing environment} 

Each JHipster configuration requires to use specific tools and pre-defined settings. Without them, the compilation, build, or execution cannot be performed. A substantial engineering effort was needed to build an integrated environment capable of deriving any JHipster configuration. 
 The concrete result is a Debian image with all tools pre-installed and pre-configured. 
 This process was based on numerous tries and errors, using some configurations. In the end, we converged on an all-inclusive environment.
 
\subsubsection{Distributing the tests}

The number of JHipster variants led us to consider strategies to scale up the execution of the testing workflow. We decided to rely on Grid'5000\footnote{\url{https://www.grid5000.fr}}, a large-scale testbed offering a large amount of computational resources~\cite{Balouek2012}. We used numerous distributed machines, each in charge of testing a subset of configurations. Small-scale experiments (e.g., on local machines) helped us to manage distribution issues in an incremental way. Distributing the computation further motivated our previous needs of testing automation and pre-set Debian images.  

\subsubsection{Opportunistic optimizations and sharing}

Each JHipster configuration requires to download numerous Java and JavaScript dependencies, which consumes bandwidth and increases JHipster variant generation time. To optimise this in a distributed setting, we downloaded all possible Maven, npm and Bower dependencies -- once and for all configurations. We eventually obtained a Maven cache of 481MB and a node\_modules (for JavaScript dependencies) of 249MB. Furthermore, we build a \textsf{Docker} variant right after the classical build (see Figure~\ref{fig:oracleWorkflow}) to derive two JHipster variants (with and without \textsf{Docker}) without restarting the whole derivation process. 
 
\subsubsection{Validation of the testing infrastructure}

A recurring reaction after a failed build was to wonder whether the failure was due to a buggy JHipster variant or an invalid assumption/configuration of our infrastructure. We extensively tried some selected configurations for which we know it should work and some for which we know it should not work. Based on some potential failures, we  reproduced them on a local machine and studied the error messages. We also used statistical methods and GitHub issues to validate some of the failures (see next Section). This co-validation, though difficult, was necessary to  gain confidence in our infrastructure. After numerous tries on our selected configurations, we launched the testing workflow for all the configurations (selected ones included).

\subsection{Human Cost}

The development of the complete derivation and testing infrastructure was achieved in about 4 months by 2 people (i.e., \textbf{8 person * month} in total). 
For each activity, we report the duration of the effort realized in the first place. Some modifications were also made in parallel to improve different parts of the solution -- we count this duration in subsequent activities. 

\paragraph{Modelling configurations.} 
The elaboration of the first major version of the feature model took us about \textbf{2 weeks} based on the analysis of the JHipster code and configurator.

\paragraph{Configuration-aware testing workflow.} 
Based on the feature model, we initiated the development of the testing workflow. We added features and testing procedures in an incremental way. The effort spanned on a period of \textbf{8 weeks}.

\paragraph{All-inclusive environment.}
The building of the Debian image was done in parallel to the testing workflow. It also lasted a period of \textbf{8 weeks} for identifying all possible tools and settings needed.

\paragraph{Distributing the computation.}
We decided to deploy on Grid'5000 at the end of November and the implementation has lasted \textbf{6 weeks}. It includes a learning phase (1 week), the optimization for caching dependencies, and the gathering of results in a central place (a CSV-like table with logs). 

\vspace{0.3cm}
\begin{tcolorbox}
\textbf{(RQ1.1)} \textit{What is the cost of engineering an infrastructure capable of automatically deriving and testing all configurations?}
The testing infrastructure is itself a configurable system and requires a substantial engineering effort (8 man-months) to cover all design, implementation and validation activities, the latter being the most difficult.
\end{tcolorbox}

\subsection{Computational Cost}
\label{sec:computationalcost}

We used a network of machines that allowed us to test all 26,256 configurations in \emph{less than a week}. Specifically, we performed a reservation of \textbf{80} machines for \textbf{4 periods} (4 nights) of \textbf{13 hours}. The analysis of 6 configurations took on average about 60 minutes.  The total CPU time of the workflow on all the configurations is \textbf{4,376 hours}. Besides CPU time, the processing of all variants also required enough free disk space. Each scaffolded Web application occupies between 400MB and 450MB, thus forming a total of \textbf{5.2 terabytes}.
 
We replicated three times our exhaustive analysis (with minor modifications of our testing procedure each time); we found similar numbers for assessing the computational cost on Grid'5000. As part of our last experiment, we observed suspicious failures for 2,325 configurations with the same error message: ``Communications link failure'', denoting network communication error (between a node and the controller for instance) on the grid. Those failures have been ignored and configurations have been re-run again afterwards to have consistent results.

\vspace{0.3cm}
\begin{tcolorbox}
\textbf{(RQ1.2)} \textit{What are the computational resources needed to test all configurations?}
Testing all configurations requires a significant amount of computational resources (4,376 hours CPU time and 5.2 terabytes of disk space). 
\end{tcolorbox}

\section{Results of the Testing Workflow Execution (RQ2.1)}
\label{sec:rq2}

The execution of the testing workflow yielded a large file comprising numerous results for each configuration. This file\footnote{Complete results are available at \url{https://github.com/xdevroey/jhipster-dataset/tree/master/v3.6.1}.} 
allows to identify failing configurations, i.e., configurations that do not compile or build. In addition, we also exploited stack traces for grouping together some failures. We present here the ratios of failures and associated faults. 

\subsection{Bugs: A Quick Inventory}
\label{subsec:inventory}

\begin{figure}[t]
  \centering
    \includegraphics[width=0.48\textwidth]{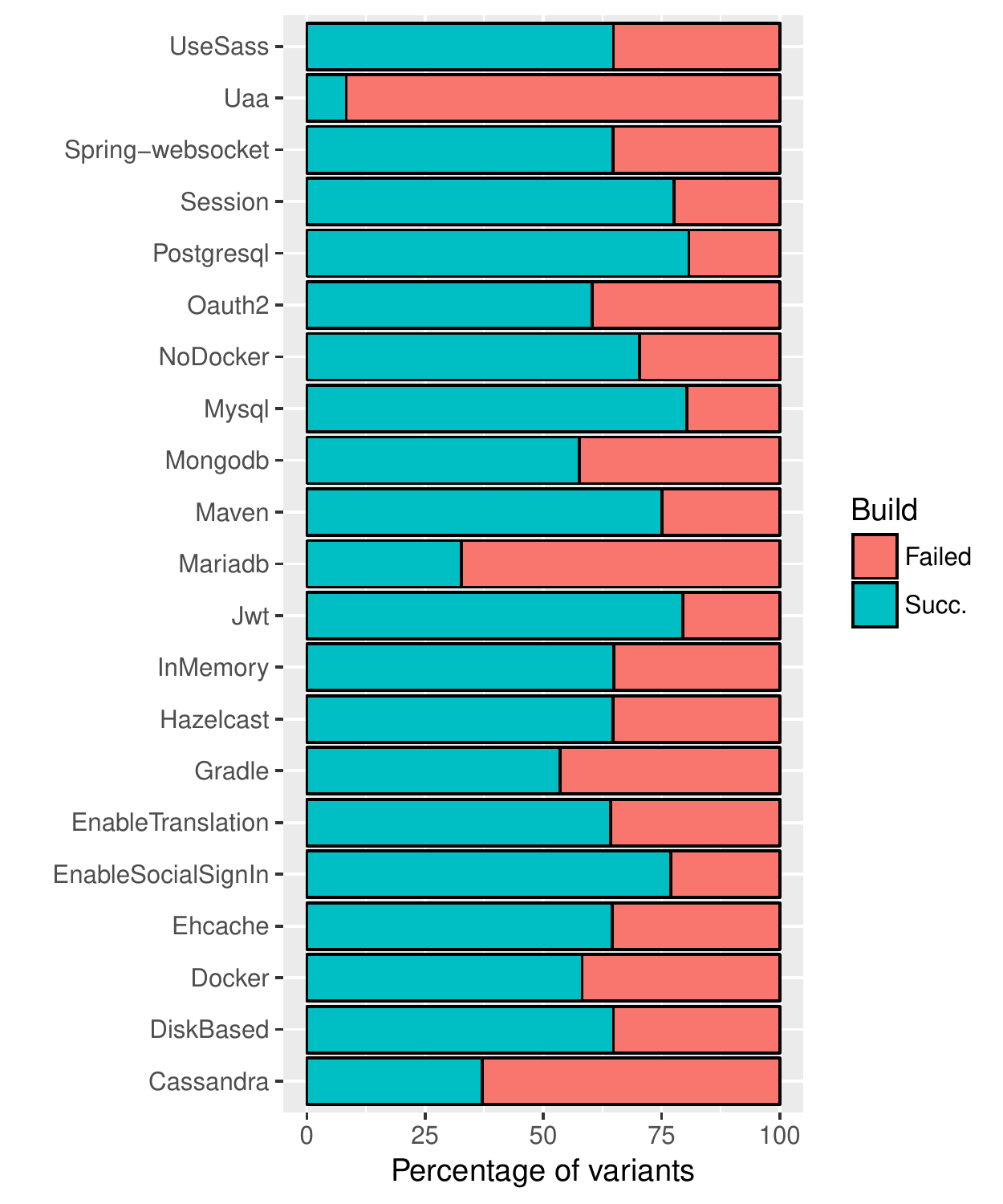}
    \caption{Proportion of build failure by feature}
    \label{fig:failurePerFeature}
\end{figure}

Out of the \textbf{26,256} configurations we tested, we found that \textbf{9,376} (\textbf{35.70}\%) failed. This failure occurred either during the compilation of the variant (\textit{Compilation} in Figure~\ref{fig:oracleWorkflow}) or during its packaging as an executable Jar file (\textit{Build Maven/Gradle} in Figure~\ref{fig:oracleWorkflow}, which includes execution of the different Java and JavaScript tests generated by JHipster), although the generation (\textit{App generation} in Figure~\ref{fig:oracleWorkflow}) was successful. We also found that some features were more concerned by failures as depicted in Figure~\ref{fig:failurePerFeature}.
Regarding the application type, for instance, \emph{microservice gateways} and \emph{microservice applications} are proportionally more impacted than \emph{monolithic applications} or \emph{UAA server} with, respectively, \textbf{58.37}\% of failures (4,184 failing microservice gateways configurations) and \textbf{58.3}\% of failures (532 failing microservice applications configurations). \emph{UAA authentication} is involved in most of the failures: \textbf{91.66}\% of \emph{UAA-based microservices} applications (4,114 configurations) fail to deploy.

\subsection{Statistical Analysis}
\label{subsec:statisticalbugs}

Previous results do not show the root causes of the configuration failures -- what features or interactions between features are involved in the failures? To investigate correlations between features and failures' results, we decided to use the Association Rule learning method~\cite{Hahsler2005f}. 
It aims at extracting relations (called \emph{rules}) between variables of large data-sets.
The Association Rule method is well suited to find the (combinations of) features leading to a failure, out of tested configurations. 

Formally and adapting the terminology of association rules, the problem can be defined as follows.
\begin{itemize}
\item let $F=\{ft_{1},ft_{2},\ldots ,ft_{n} , bs\}$ be a set of $n$ features ($ft_i$) plus the status of the build ($bs$), i.e., build failed or not; 
\item let $C=\{c_{1},c_{2},\ldots ,c_{m}\}$ be a set of $m$ configurations. 
\end{itemize}
Each configuration in $C$ has a unique identifier and contains a subset of the features in $F$ and the status of its build. A rule is defined as an implication of the form: $X \Rightarrow Y$, where $X,Y \subseteq F$.
The outputs of the method are a set of rules, each constituted by: 
\begin{itemize}
\item $X$ the \emph{left-hand side (LHS)} or antecedent of the rule;
\item $Y$ the \emph{right-hand side (RHS)} or consequent of the rule.
\end{itemize}

\begin{table*}[t]
\centering
\caption{An example of JHipster data (feature values and build status for each configuration). We want to extract association rules stating which combinations of feature values lead to a build failure (e.g., gradle).}
\label{tab:jhipsterAR}
\begin{tabular}
{c | c c c c c | c}
	\hline 
    \textbf{Conf.} & \textbf{gradle} & \textbf{mariadb}& \textbf{enableSocialSignIn} & \textbf{websocket} & \textbf{...} & \textbf{build failure}\\ \hline
    1 & 1 & 0 & 0 & 0 & ... & 0 \\
    2 & 0 & 1 & 0 & 0 & ... & 0 \\
    3 & 0 & 0 & 1 & 1 & ... & 0 \\
    4 & 1 & 1 & 0 & 0 & ... & 1 \\
    5 & 1 & 0 & 0 & 0 & ... & 0 \\
    6 & 1 & 1 & 0 & 0 & ... & 1 \\
    ... & ... & ... & ... & ... & ... & ... \\
\end{tabular}
\end{table*}

For our problem, we consider that $Y$ is a single target: the status of the build. For example, we want to understand what combinations of features lead to a failure, either during the compilation or the build process. 
To illustrate the method, let us take a small example (see Table~\ref{tab:jhipsterAR}).
The set of features is $F=\{$ \textit{mariadb}, \textit{gradle}, \textit{enable\-Social\-Sign\-In}, \textit{web\-so\-cket}, \textit{failure} $\}$ and in the table is shown a small database containing the configurations, where, in each entry, the value 1 means the presence of the feature in the corresponding configuration, and the value 0 represents the absence of a feature in that configuration. In Figure~\ref{tab:jhipsterAR}, when \textit{build failure} has the value 1 (resp. 0), it means the build failed (resp. succeeded). 
An example rule could be: 
$$\{\mathrm {mariadb,graddle} \} \Rightarrow \{\mathrm {build\ failure} \}$$
Meaning that if \textit{mariadb} and \textit{gradle} are activated, configurations will not build.


As there are many possible rules, some well-known measures are typically used to select the most interesting ones. 
In particular, we are interested in the \emph{support}, the proportion of configurations where LHS holds and the \emph{confidence}, the proportion of configurations where both LHS and RHS hold.
In our example and for the rule $\{\mathrm {mariadb,graddle} \}\Rightarrow \{\mathrm {build\ failure} \}$, the support is $2/6$ while the confidence is $1$. 

\begin{table*}[t]
\centering
\caption{Association rules involving compilation and build failures}
\label{tab:arules}
\begin{tabular}
{l p{4.2cm}p{2.0cm} r r p{1.35cm}p{3.5cm}}
	\hline 
    \textbf{Id} & \textbf{Left-hand side} & \textbf{Right-hand side} & \textbf{Support}& \textbf{Conf.} & \textbf{GitHub Issue}& \textbf{Report/Correction date}\\ \hline
    \textsc{MoSo}
	& DatabaseType=``mongodb",\newline
EnableSocialSignIn=true
    & Compile=KO 
    & 0.488 \%
    & 1
    & 4037
    & 27 Aug 2016 (report and fix for milestone 3.7.0)\\ 
    \textsc{MaGr}
	& prodDatabaseType=``mariadb",\newline
 buildTool=``gradle"
    & Build=KO 
    & 16.179 \%
    & 1
    & 4222
    & 27 Sep 2016 (report and fix for milestone 3.9.0) \\ 
    \textsc{UaDo}
	& Docker=true,\newline
authenticationType=``uaa" 
    & Build=KO 
    & 6.825 \%
    & 1
    & UAA is in Beta
    & Not corrected \\ 
    \textsc{OASQL}
	& authenticationType=``uaa",\newline 
hibernateCache=``no"
    & Build=KO 
    & 2.438 \%
    & 1
    & 4225
    & 28 Sep 2016 (report and fix for milestone 3.9.0) \\ 
    \textsc{UaEh}    
	& authenticationType=``uaa",\newline
hibernateCache=``ehcache"
    & Build=KO 
    & 2.194 \%
    & 1
    & 4225
    & 28 Sep 2016 (report and fix for milestone 3.9.0) \\ 
    \textsc{MaDo}
	& prodDatabaseType=``mariadb",\newline
applicationType=``monolith",\newline
searchEngine=``false",\newline
Docker=``true"
    & Build=KO
    & 5.590\%
    & 1
    & 4543
    & 24 Nov 2016 (report and fix for milestone 3.12.0) \\ \hline
\end{tabular}
\end{table*}

Table~\ref{tab:arules} gives some examples of the rules we have been able to extract. We parametrized the method as follows. First, we restrained ourselves to rules where the RHS was a failure: either \textit{Build=KO} (build failed) or \textit{Compile=KO} (compilation failed). Second, we fixed the confidence to 1: if a rule has a confidence below 1 then it is not asserted in all configurations where the LHS expression holds -- the failure does not occur in all cases. Third, we lowered the support in order to catch all failures, even those afflicting smaller proportion of the configurations. For instance, only 224 configurations fail due to a compilation error; in spite of a low support, we can still extract rules for which the RHS is \textit{Compile=KO}. We computed redundant rules using facilities of the R package \textsf{arules}.\footnote{\url{https://cran.r-project.org/web/packages/arules/}} As some association rules can contain already known constraints of the feature model, we ignored some of them. 

We first considered association rules for which the size of the LHS is either 1, 2 or 3. We extracted 5 different rules involving two features (see Table~\ref{tab:arules}). We found no rule involving 1 or 3 features. We decided to examine the 200 association rules for which the LHS is of size 4. We found out a sixth association rule that incidentally corresponds to one of the first failures we encountered in the early stages of this study. 

\begin{figure}[t]
  \centering
    \includegraphics[width=0.49\textwidth]{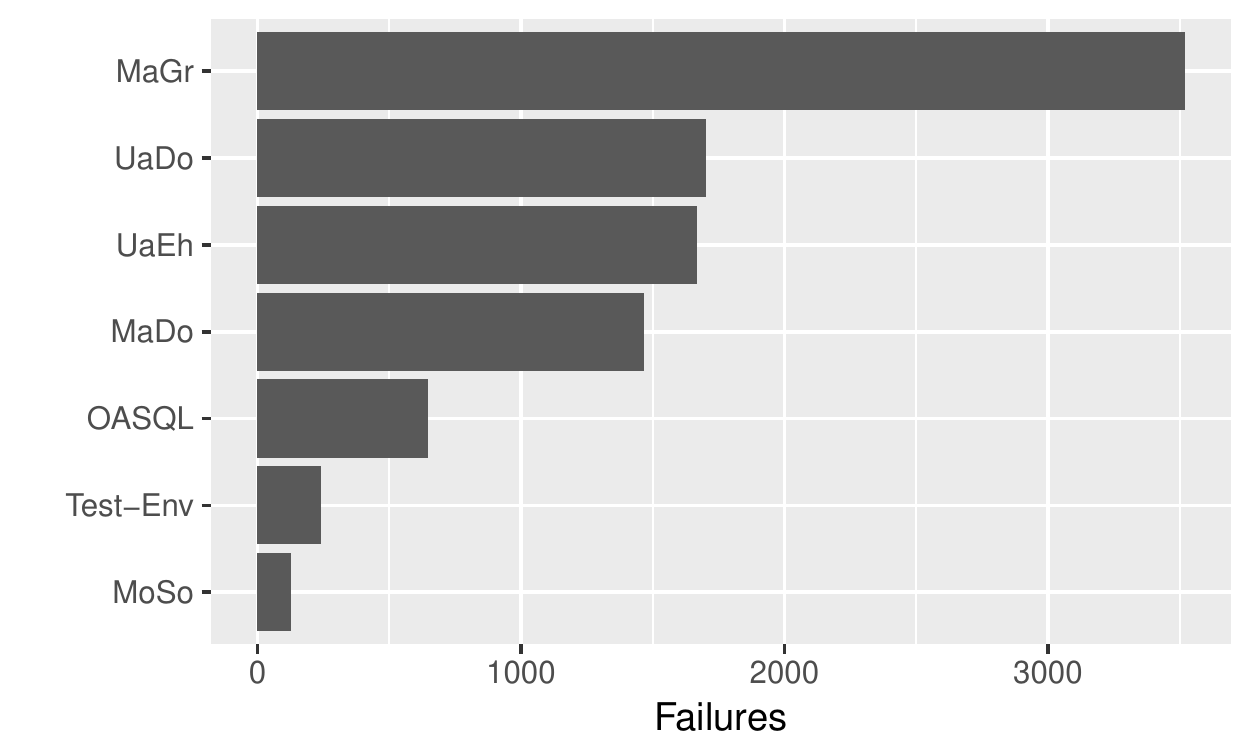}
    \caption{Proportion of failures by fault described in Table \ref{tab:arules}.}
    \label{fig:resultsPercentage}
\end{figure}

Table~\ref{tab:arules} shows that there is only one rule with the RHS being \emph{Compile=KO}. According to this rule, all configurations in which the database is \textsf{MongoDB} and \textsf{social login} feature is enabled (128 configurations) fail to compile. The other 5 rules are related to a build failure. Figure~\ref{fig:resultsPercentage} reports on the proportion of failed configurations that include the LHS of each association rule. Such LHS can be seen as a feature interaction fault that causes failures. For example, the combination of \textsf{MariaDB} and \textsf{Gradle} explains \emph{37\%} of failed configurations (or 13\% of all configurations). We conclude that six feature interaction faults explain 99.1\% of the failures. 

\subsection{Qualitative Analysis}
\label{subsec:qualitativebugs}

We now characterize the 6 important faults, caused by the interactions of several features (between 2 features and 4 features). Table~\ref{tab:arules} gives the support, confidence for each association rule. We also confirm each fault by giving the GitHub issue and date of fix. 

\paragraph{MariaDB with Docker.}
This fault is the only one caused by the interaction of 4 features: it concerns \emph{monolithic} web-applications relying on \textsf{MariaDB} as production data\-base, where the search-engine (\textsf{ElasticSearch}) is disabled and built with \textsf{Docker}. These variants amount to \textbf{1,468} configurations and the root cause of this bug lies in the template file \textit{src/main/docker/\_app.yml} where a condition (\emph{if prodDB = MariaDB}) is missing.

\paragraph{MariaDB using Gradle.}
This second fault concerns variants relying on \textsf{Gradle} as build tool and \textsf{MariaDB} as the database (\textbf{3,519} configurations). It is caused by a missing dependency in template file \textit{server/template/gradle/\_liquibase.gradle}.

\paragraph{UAA authentication with Docker.}
The third fault occurs in \textsf{Microservice Gateways} or \textsf{Microservice applications} using an \textsf{UAA} server as authentication mechanism (\textbf{1,703} Web apps). This bug is encountered at build time, with \textsf{Docker}, and it is due to the absence of UAA server Docker image. It is a known issue, but it has not been corrected yet, UAA servers are still in beta versions.

\paragraph{UAA authentication with Ehcache as Hibernate second level cache.}
This fourth fault concerns \textsf{Microservice Gateways} and \textsf{Microservice applications}, using a \textsf{UAA} authentication mechanism. When deploying manually (i.e., with \textsf{Maven} or \textsf{Gradle}), the web application is unable to reach the deployed UAA instance. This bug seems to be related to the selection of Hibernate cache and impacts \textbf{1,667} configurations.

\paragraph{OAuth2 authentication with SQL database.}
This defect is faced \textbf{649} times, when trying to deploy a web-app, using an SQL database (\textsf{MySQL}, \textsf{PostgreSQL} or \textsf{MariaDB}) and an \textsf{OAuth2} authentication, with \textsf{Docker}. It was reported on August 20th, 2016 but the JHipster team was unable to reproduce it on their end.

\paragraph{Social Login with MongoDB.}
This sixth fault is the only one occurring at compile time. Combining \textsf{MongoDB} and \textsf{social login} leads to \textbf{128} configurations that fail. The source of this issue is a missing import in class \textit{SocialUserConnection.java}. This import is not in a conditional compilation condition in the template file while it should be.

\paragraph{Testing infrastructure.} \label{sec:infraissues}
We have not found a common fault for the remaining \textbf{242} configurations that fail. We came to this conclusion after a thorough and manual investigation of all logs.\footnote{Such configurations are tagged by ``ISSUE:env'' in the column ``bug'' of the JHipster results CSV file available online \url{https://github.com/xdevroey/jhipster-dataset}.} We noticed that, despite our validation effort with the infrastructure (see \textbf{RQ1}), the observed failures are caused by the testing tools and environment. Specifically, the causes of the failures can be categorized in two groups: \textit{(i)} several network access issues in the grid that can affect the testing workflow at any stage and \textit{(ii)} several unidentified errors in the configuration of building tools (\texttt{gulp} in our case). 
\paragraph{Feature interaction strength.} 
Our findings show that only two features are involved in five (out of six) faults, and four features are involved in the last fault.
The prominence of 2-wise interactions is also found in other studies. Abal \etal report that, for the Linux bugs they have qualitatively examined, more than a half (22/43) are attributed to 2-wise interactions \cite{DBLP:journals/tosem/AbalMSBRW18}.  Yet, for different interaction strengths, there is no common trend: we do not have 3-wise interactions while this is second most common case in Linux, we did not find any fault caused by one feature only.     


\vspace{0.3cm}
\begin{tcolorbox}
\textbf{(RQ2.1)} \textit{How many and what kinds of failures/faults can be found in all configurations?}
Exhaustive testing shows that almost 36\% of the configurations fail. Our analysis identifies 6 interaction faults as the root cause for this high percentage. 
\end{tcolorbox}

\section{Sampling Techniques Comparison (RQ2.2)} 
\label{subsec:jhipstersampling}

In this section, we first discuss the sampling strategy used by the JHipster team. We then use our dataset to make a ground truth comparison of six state-of-the-art sampling techniques.

\subsection{JHipster Team Sampling Strategy}

The JHipster team uses a sample of 12 representative configurations for the version 3.6.1, to test their generator (see Section \ref{subsec:jhipsterTesting} for further explanations on how these were sampled). During a period of several weeks, the testing configurations have been used at \emph{each commit} (see also Section~\ref{subsec:jhipsterTesting}). These configurations fail to reveal any problem, i.e., the Web-applications corresponding to the configurations successfully compiled, build and run. We assessed these configurations with our own testing infrastructure and came to the same observation. We thus conclude that this sample was not effective to reveal any defect.

\begin{table*}[t]
\centering
\caption{Efficiency of different sampling techniques (bold values denote the highest efficiencies)}
\label{tab:comparison}
\begin{tabular}
{l r r r r r}
	\hline \textbf{Sampling technique} & \textbf{Sample size} & \textbf{Failures} ($\sigma$)& \textbf{Failures eff.}& \textbf{Faults} ($\sigma$)& \textbf{Fault eff.}\\ 
    \hline 
    1-wise & 8 & 2.000 (N.A.) & 25.00\% & 2.000 (N.A.) & 25.00\% \\ 
    Random(8) & 8 & 2.857 (1.313) & 35.71\% & 2.180 (0.978) & 27.25\% \\ 
    PLEDGE(8) & 8 & 3.160 (1.230) & 39.50\% & 2.140 (0.825) & 26.75\% \\ 
    Random(12) & 12 & 4.285 (1.790) & 35.71\% & 2.700 (1.040) & 22.5\% \\ 
    PLEDGE(12) & 12 & 4.920 (1.230) & 41.00\% & 2.820 (0.909) & 23.50\% \\ 
    2-wise & 41 & 14.000 (N.A.) & 34.15\% & 5.000 (N.A.) & 12.20\% \\ 
    Random(41) & 41 & 14.641 (3.182) & 35.71\% & 4.490 (0.718) & 10.95\% \\ 
    PLEDGE(41) & 41 & 17.640 (2.500) & 43.02\% & 4.700 (0.831) & 11.46\% \\ 
    3-wise & 126 & 52.000 (N.A.) & 41.27\% & 6.000 (N.A.) & 4.76\% \\ 
    Random(126) & 126 & 44.995 (4.911) & 35.71\% & 5.280 (0.533) & 4.19\% \\ 
    PLEDGE(126) & 126 & 49.080 (11.581) & 38.95\% & 4.660 (0.698) & 3.70\% \\ 
    4-wise & 374 & 161.000 (N.A.) & \textbf{43.05\%} & 6.000 (N.A.) & 1.60\% \\ 
    Random(374) & 374 & 133.555 (8.406) & 35.71\% & 5.580 (0.496) & 1.49\% \\ 
    PLEDGE(374) & 374 & 139.200 (31.797) & 37.17\% & 4.620 (1.181) & 1.24\% \\ 
	Most-enabled-disabled & 2 & 0.683 (0.622) & 34.15\% & 0.670 (0.614) & \textbf{33.50\%} \\ 
    All-most-enabled-disabled & 574 & 190.000 (N.A.) & 33.10\% & 2.000 (N.A.) & 0.35\% \\ 
    One-disabled & 34 & 7.699 (2.204) & 0.23\% & 2.398 (0.878) & 0.07\% \\ 
    All-one-disabled & 922 & 253.000 (N.A.) & 27.44\% & 5.000 (N.A.) & 0.54\% \\ 
    One-enabled & 34 & 12.508 (2.660) & 0.37\% & 3.147 (0.698) & 0.09\% \\ 
    All-one-enabled & 2,340 & 872.000 (N.A.) & 37.26\% & 6.000 (N.A.) & 0.26\% \\ 
    ALL & 26,256 & 9,376.000 (N.A.) & 35.71\% & 6.000 (N.A.) & 0.02\% \\ 
    \hline
\end{tabular}
\end{table*}

\subsection{Comparison of Sampling Techniques}
\label{subsec:twisesampling}

As testing all configurations is very costly (see \textbf{RQ1}), sampling techniques remain of interest. We would like to find as many failures and faults as possible with a minimum of configurations in the sampling. For each failure, we associate a fault through the automatic analysis of features involved in the failed configuration (see previous subsections). 

We address \textbf{RQ2.2} with  numerous sampling techniques considered in the literature~\cite{Perrouin:2010,Johansen2012,Abal:2014,MKRGA:ICSE16}. For each technique, we report on the number of failures and faults. 

\subsubsection{Sampling techniques}

\paragraph{\textit{t}-wise sampling.} We selected 4 variations of the \textit{t}-wise criteria: \textbf{1-wise}, \textbf{2-wise}, \textbf{3-wise} and \textbf{4-wise}. We generate the samples with \textit{SPLCAT}~\cite{Johansen2012}, which has the advantage of being deterministic: for one given feature model, it will always provide the same sample. 
The 4 variations yield samples of respectively \textbf{8}, \textbf{41}, \textbf{126} and \textbf{374} configurations. \textbf{1-wise} only finds \textbf{2} faults; \textbf{2-wise} discovers \textbf{5 out of 6 faults}; \textbf{3-wise} and \textbf{4-wise} find \textbf{all} of them. It has to be noted that the discovery of a 4-wise interaction fault with a 3-wise setting is a `collateral' effect \cite{Petke:2013:EEF:2491411.2491436}, since any sample covering completely t-way interactions also yields an incomplete coverage of higher-order interactions.  

\paragraph{One-disabled sampling.} Using \textbf{one-disabled} sampling algorithm, we extract configurations in which all features are activated but one. 
To overcome any bias in selecting the first valid configuration, as suggested by Medeiros \etal~\cite{MKRGA:ICSE16}, we applied a random selection instead. We therefore select a valid random configuration for each disabled feature (called \textbf{one-disabled} in our results) and repeat experiments 1,000 times to get significant results. 
This gives us a sample of \textbf{34 configurations} which detects on average \textbf{2.4 faults} out of 6.

Additionally, we also retain \textbf{all-one-disabled} configurations (i.e., all valid configurations where one feature is disabled and the other are enabled). The all-one-disabled sampling yields a total sample of \textbf{922 configurations} that identifies \textbf{all faults but one}. 

\paragraph{One-enabled and most-enabled-disabled sampling.} In the same way, we implemented sampling algorithms covering the \textbf{one-enabled} and \textbf{most-enabled-disabled} criteria \cite{Abal:2014,MKRGA:ICSE16}. 
%
As for one-disabled, we choose to randomly select valid configurations instead of taking the first one returned by the solver. Repeating the experiment  1,000 times: one-enabled extracts a sample of \textbf{34 configurations} which detects \textbf{3.15 faults} on average; and most-enabled-di\-sa\-bled gives a sample of \textbf{2 configurations} that detects \textbf{0.67 faults} on average.
Considering all valid configurations, \textbf{all-one-enabled} extracts a sample of \textbf{2,340 configurations} and identifies all the \textbf{6 faults}. \textbf{All-most-enabled-disabled} gives a sample of \textbf{574 configurations} that identifies \textbf{2 faults} out of 6.

\paragraph{Dissimilarity sampling.} We also considered \emph{dissimilarity} testing for software product lines \cite{Henard2014a,Al-Hajjaji2016} using \textbf{PLE\-DGE} \cite{Henard2013PLEDGE}. 
We retained this technique because it can afford any testing budget (sample size and generation time).  For each sample size, we report the average failures and faults for 100 PLEDGE executions with the greedy method in 60 secs~\cite{Henard2013PLEDGE}. 
We selected (respectively) \textbf{8}, \textbf{12}, \textbf{41}, \textbf{126} and \textbf{374} configurations, finding (respectively) \textbf{2.14}, \textbf{2.82}, \textbf{4.70}, \textbf{4.66} and \textbf{4.60} faults out of 6.

\paragraph{Random sampling.} Finally, we considered \textbf{random} samples from size 1 to 2,500. The random samples exhibit, by construction, 35.71\% of failures on average (the same percentage that is in the whole dataset). To compute the number of unique faults, we simulated 100  random selections. We find, on average, respectively \textbf{2.18}, \textbf{2.7}, \textbf{4.49}, \textbf{5.28} and \textbf{5.58} faults for respectively \textbf{8}, \textbf{12}, \textbf{41}, \textbf{126} and \textbf{374} configurations. 

\subsubsection{Fault and failure efficiency}

We consider two main metrics to compare the efficiency of sampling techniques to find faults and failures w.r.t the sample size. 
\emph{Failure efficiency} is the ratio of \textit{failures to sample size}. \emph{Fault efficiency} is the ratio of \textit{faults to sample size}. For both metrics, a high efficiency is desirable since it denotes a small sample with either a high failure or fault detection capability.

\begin{figure*}[t]
  \centering
  \subfigure[Failures found by sampling techniques]{
    \includegraphics[height=7cm,trim={0 0 4cm 0},clip]{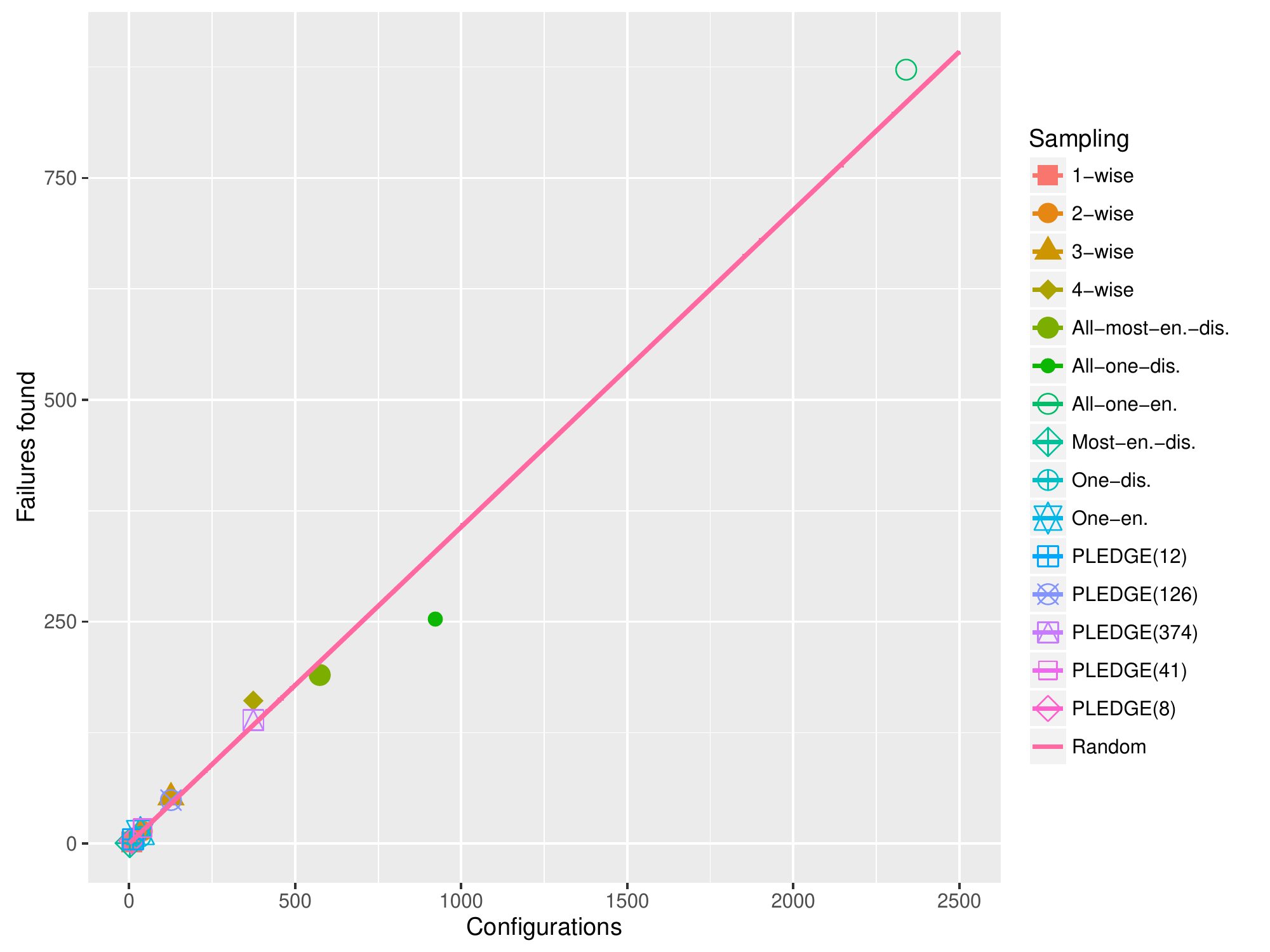}
    \label{fig:samplingComparisonFailures}
  }
  \subfigure[Faults found by sampling techniques]{
    \includegraphics[height=7cm]{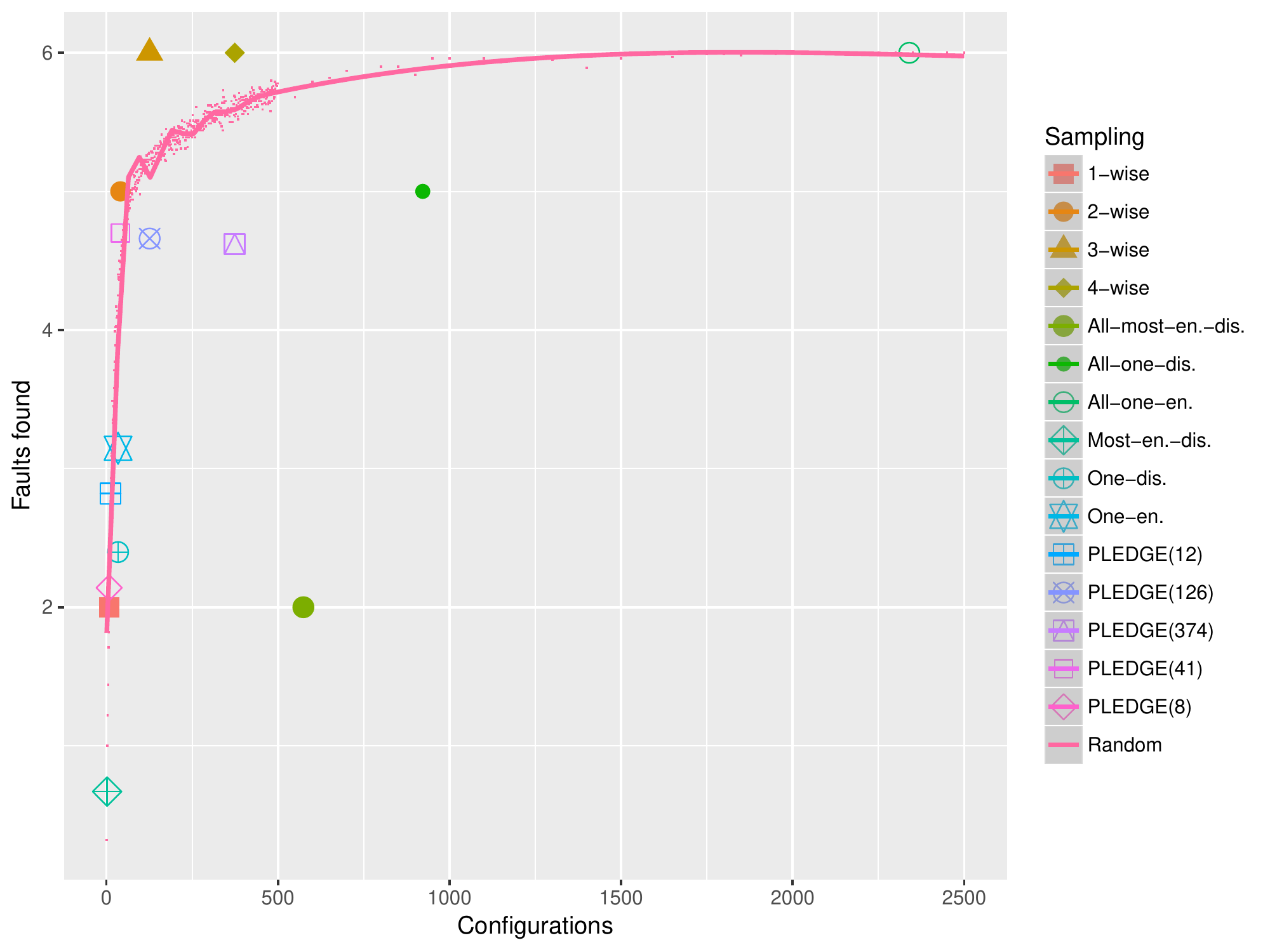}
    \label{fig:samplingComparisonBugs}
  }
  \caption{Defects found by sampling techniques}
\end{figure*}

The results are summarized in Table~\ref{tab:comparison}. We present in Figure~\ref{fig:samplingComparisonFailures} (respectively, Figure~\ref{fig:samplingComparisonBugs}) the evolution of \textit{failures} (respectively, \textit{faults}) w.r.t. the size of random samples. To ease comparison, we place reference points corresponding to results of other sampling techniques. A first observation is that random is a strong baseline for both failures and faults. 2-wise or 3-wise sampling techniques are slightly more efficient to identify faults than random. On the contrary, all-one-enabled, one-enabled, all-one-disabled, one-disabled and ~all-most-enabled-disabled identify less faults than random samples of the same size. 
Most-enabled-disabled is efficient on average to detect faults (33.5\% on average) but requires to be ``lucky". In particular, the first configurations returned by the solver (as done in~\cite{MKRGA:ICSE16}) discovered 0 fault. 
This shows the sensitivity of the selection strategy amongst valid configurations matching the most-enabled-disabled criterion. Based on our experience, we recommend researchers the use of a random strategy instead of picking the first configurations when assessing one-disabled, one-enabled, and most-enabled-disabled.  
 
PLEDGE is superior to random for small sample sizes. The significant difference between 2-wise and 3-wise is explained by the sample size: although the latter finds all the bugs (one more than 2-wise) its sample size is triple (126 configurations against 41 for 2-wise). In general, a relatively small sample is sufficient to quickly identify the 5 or 6 most important faults -- there is no need to cover the whole configuration space. 

A second observation is that there is no correlation between failure efficiency and fault efficiency. For example, all-one-enabled has a failure efficiency of 37.26\% (better than random and many techniques) but is one of the worst techniques in terms of fault rate due of its high sample size. In addition, some techniques, like all-most-enable-disabled, can find numerous failures that in fact correspond to the same fault. 

\subsubsection{Discussion}

Our results show that the choice of a metric (failure-detection or fault-detection capability) can largely influence the choice of a sampling technique. Our initial assumption was that the detection of one failure leads to the finding of the associated fault. 
The GitHub reports and our qualitative analysis show that it is indeed the case in JHipster: contributors can easily find the root causes based on a \emph{manual} analysis of a configuration failure. 
For other cases, finding the faulty features or feature interactions can be much more tricky. In such contexts, investigating \emph{many failures} and using statistical methods (such as association rules) can be helpful to determine the faulty features and their undesired interactions. As a result, the ability of finding failures may be more important than in JHipster case. A trade-off between failure and fault efficiency can certainly be considered when choosing the sampling technique. 

\vspace{0.3cm}
\begin{tcolorbox}
\textbf{(RQ2.2)} \textit{How effective are sampling techniques comparatively?}
To summarise: 
\begin{inparaenum}[(i)]
\item random is a strong baseline for failures and faults;
\item 2-wise and 3-wise sampling are slightly more efficient to find faults than random;
\item most-enabled-disabled is efficient on average to detect faults but requires to be lucky;
\item dissimilarity is superior to random for small sample sizes;
\item a small sample is sufficient to identify most important faults, there is no need to cover the whole configuration space; and
\item there is no correlation between failure and fault efficiencies.
\end{inparaenum}
\end{tcolorbox}

%% file: parts_otherstudies.tex
\section{Comparison with Other Studies (RQ2.3)}
\label{sec:otherstudied}

This section presents a literature review of case studies of configuration sampling approaches to test variability intensive systems. 
 Specifically, we aim to compare our findings with state-of-the-art results: 
\emph{Are sampling techniques as effective in other case studies? Do our results confirm or contradict findings in other settings?}
This question is important for (1) practitioners in charge of establishing a suitable strategy for testing
their systems; (ii) researchers interested in building evidence-based theories
or tools for testing configurable systems.

We first present our selection protocol of relevant papers and an overview of the selected ones. We then confront and discuss our findings from Section \ref{subsec:twisesampling} w.r.t. those studies.

\subsection{Studies Selection Protocol}

We consider the following criteria to select existing studies:
\begin{inparaenum}[(i)]
\item configuration sampling approaches are evaluated regarding defects detection capabilities;
\item evaluation has been performed on an industrial size (open source or not) system (i.e., we discard toy examples); and
\item evaluation has been performed on the system (possibility to analyse the source code and/or to run the variants to reproduce bugs and failures). We thus discard evaluations that are solely based on feature models such as \cite{Perrouin2011}.
\end{inparaenum}

We looked for studies in previous literature reviews on product line testing \cite{Engstrom2011,DaMotaSilveiraNeto2011,Machado2014}. They are a common way to give an overview of a research field: e.g., they organise studies according to topic(s) and validation level (for instance, from \cite{Machado2014}: no evidence, toy example, opinions or observations, academic studies, industrial studies, or industrial practices). 
Before 2014 (i.e., before the publication of the systematic literature review from Machado \etal~\cite{Machado2014}), empirical evaluations of configurations sampling approaches are focused on their capability to select a sampling satisfying t-wise criteria in a reasonable amount of time or with fewer configurations \cite{Ensan2012,Henard2013a,Hervieu2011,Johansen2011,Johansen2012,kim2013splat,Lochau2011,Perrouin2010a,Perrouin2011}. We discarded them as they do not match our selection criteria.

To select relevant studies without performing a full systematic literature survey, we applied a forward and backward snowballing search~\cite{Jalali2012}. Snowballing is particularity relevant in our case, given the diversity of terms used in the research literature (product line, configurable systems, etc.) and our goal to compare more than one sampling approach. Searching and filtering studies from literature databases would require a large amount of work with few guarantees on the quality of the result. We started the search with two empirical studies known by the authors of the paper: Medeiros et~al.~\cite{MKRGA:ICSE16} and S\'anchez et~al.~\cite{Sanchez2017}. 
Those studies are from two different research sub-communities on variability-intensive system testing, configurable systems research (Medeiros et~al.~\cite{MKRGA:ICSE16}) and software product line research (S\'anchez \etal~\cite{Sanchez2017}), which mitigates the risk of missing studies of interest. Eventually, we collected 5 studies presented in Table \ref{tab:litComparison} and  we discuss them below.

\subsection{Selected studies}

\begin{table*}[t]
\centering
\caption{Evaluation of configuration sampling techniques in the literature}
\label{tab:litComparison}
\begin{small}
\rowcolors{2}{}{gray!15}
\begin{tabular}{p{1.5cm} p{7cm} p{7.6cm}}
\hline 
\textbf{Reference} & \textbf{Samplings} & \textbf{Validation} \\
\hline 
Medeiros et~al.~\cite{MKRGA:ICSE16} 
    & Statement-coverage, one-enabled, one-disabled, most-enabled-disabled, random, pair-wise, three-wise, four-wise, five-wise, six-wise
    & 135 configuration-related faults in 24 open-source C (\texttt{\#ifdef}) configurable systems \\
S\'anchez et~al.~\cite{Sanchez2017} 
    & Pairwise
    & Drupal (PHP modules based Web content management system) \\
Parejo et~al.~\cite{Parejo2016} 
    & Multi-objective
    & Drupal (PHP modules based Web content management system) \\
Souto et~al.~\cite{Souto2017} 
    & random, one-enabled, one-disabled, most-enabled-disabled and pairwise computed from SPLat \cite{kim2013splat}
    &  8 small SPLs + GCC (50 most used options). Samplings' sizes and number of failures were considered. \\
Apel et~al.~\cite{Apel2013a} 
    & one-wise, pairwise, three-wise compared to a family-based stategy and enumeration of all products
    & 3 configurable systems written in C and 3 in JAVA \\
\hline 
\end{tabular}  
\end{small}
\end{table*}

Medeiros \etal~\cite{MKRGA:ICSE16} compared 10 sampling algorithms using a a corpus of 135 known configuration-related faults from 24 open-source C systems. Like for JHipster, the systems use conditional compilation (\texttt{\#ifdef}) to implement variability. 

S\'anchez \etal~\cite{Sanchez2017} studied Drupal\footnote{\url{https://www.drupal.org}}, a PHP web content management, to assess test case prioritization, based on functional and non-func\-tio\-nal data extracted from Drupal's Git repository. 
 S\'anchez \etal assimilate a Drupal module to a feature and performed extensive analysis of Drupal's Git repository and issue tracking system to identify faults and other functional and non-functional attributes (e.g., feature size, feature cyclomatic complexity, number of test for each feature, feature popularity, etc.). 
 S\'anchez \etal consider 2-wise to sample the initial set of configurations to prioritize. 

Parejo \etal~\cite{Parejo2016} extend S\'anchez \etal~\cite{Sanchez2017} work by defining multi-objectives test case selection and prioritization. The algorithm starts with a set of configurations samples, each satisfying 2-wise coverage, and evolves them in order to produce one prioritized configuration sample that maximize the defined criteria. Since new configurations may be injected during the evolution, the algorithm does not only prioritize (unlike for S\'anchez \etal~\cite{Sanchez2017}), but modifies the samples. Objectives are functional (e.g., dissimilarity amongst configurations, pairwise coverage, cyclomatic complexity, variability coverage, etc.) and non-functional (e.g., number of changes in the features, feature size, number of faults in previous version, etc.). 
 
Souto \etal \cite{Souto2017} explore the tension between soundness and efficiency for configurable systems. They do so by extending SPLat \cite{kim2013splat}, a variability-aware dynamic technique that, for each test, monitors configuration variables accessed during test execution and change their values to run the test on new configurations, stopping either when no new configurations can be dynamically explored or a certain threshold is met. The extension, called S-SPLAT (Sampling with SPLat) uses dynamically explored configurations in the goal of meeting a particular sampling criterion. Six heuristics are considered: random, one-enabled, one-disabled, most-enabled-disabled and 2-wise in addition of the original SPLat technique. Experiments carried out on eight SPLs and the GCC compiler considered efficiency (number of configurations explored per test) and efficacy (number of failures detected). 

Apel \etal investigated the relevance of family-based model checking (thanks to the SPLVerifier tool chain developed by the authors) with respect to t-wise sampling strategies and analysis all products independently on C and JAVA configurable systems \cite{Apel2013a}. The metrics used were the analysis time and the number of faults founds statically.  While both the family-based strategy and the product-based (all-products) strategies covered all the faults (by construction), there is a clear advantage in favour of the former with respect to execution time. Sampling strategies were in between these two extremes in which 3-wise appears to the best compromise.

\subsection{Comparison of Findings}

\subsubsection{Sampling effectiveness} 

Souto \etal reported that one-disabled and a combination involving one-enabled, one-disabled, most-ena\-bled-di\-sa\-bled as well as pairwise appeared to be good compromises for detecting failures. 
Regarding GCC, one-enabled and most-e\-na\-bled-di\-sa\-bled were the most appropriate choices.  
On the one hand, our findings concur with their results: 
\begin{itemize}
\item 2-wise is indeed one of the most effective sampling technique, capable of identifying numerous failures and the 5 most important faults; 
\item most-enabled-disabled is also efficient to detect failures (34.15\%) and faults (33.5\% on average). 
\end{itemize}
On the other hand, our results also show some differences:
\begin{itemize}
\item one-enabled and one-disabled perform very poorly in our case, requiring a substantial number of configurations to find either failures or faults; 
\item despite a high fault efficiency, most-enabled-disabled is only able to capture 0.670 faults on average, thus missing important faults.  
\end{itemize}

Medeiros \etal's results show that most-enabled-disabled offers a good compromise for faults finding ability. 
 On the one hand, our findings concur with their results -- most-enabled-disabled is indeed efficient to detect faults (33.5\% on average). 
On the other hand, our experiments reveal an important insight. Amongst valid configurations matching the most-enabled-disabled criterion, some may not reveal any fault. It is the case in our study: the first configurations returned by the solver (as done in~\cite{MKRGA:ICSE16}) discovered 0 fault. 
For a fair comparison, we thus caution researchers to use a random strategy instead of picking the first configurations when assessing most-enabled-disabled.  
Besides Medeiros \etal reported that 2-wise is an efficient sampling technique. We concur with this result. 

Putting all together our findings and results of Souto \etal and Medeiros \etal, we can recommend the following: \emph{most-enabled-disabled is an interesting candidate to initiate the testing of configurations; 2-wise can be used to complement and continue the effort in order to find further faults.} 

S\'anchez \etal have chosen 2-wise (using the ICPL algorithm~\cite{Johansen2012} and CASA~\cite{garvin2011evaluating}) to sample the initial set of configurations of Drupal. 
Their results suggest that 2-wise is an efficient sampling technique (though we ignore how pairwise competes with other sampling strategies). 
As a follow up of their work on Drupal, Parejo \etal concluded that a combination of 2-wise and other objectives (e.g., based on non-functional properties) is usually effective.

In our case, 2-wise is more efficient to identify faults than random, offering a good balance
between sample size and fault detection.
Overall \emph{our findings on 2-wise concur with the results of S\'anchez \etal and Parejo \etal}


Apel \etal considered various sampling strategies. 3-wise appears to be the best compromise, offering a good balance between execution time and fault-detection ability. 
In our case, 3-wise is slightly more efficient to identify faults than random and can identify the 6 most important faults. 
However, the important size of 3-wise sampling (126 configurations) degrades its fault efficiency. 
In particular, 2-wise offers a better trade-off between sampling size and fault detection -- it only misses one fault despite having divided the number of configurations to assess by three (41). 

Overall, we concur with the findings of Apel \etal There is no need to consider all configurations and \emph{t-wise samplings offer a good trade-off between sampling size and fault detection}. 
The value of $t$ (e.g., 2-wise or 3-wise) and the underlying trade-off should then be debated along the specificities of the software engineering context -- it is the role of the next section in which we gather qualitative insights from JHipster community.



\subsubsection{Failure vs Fault}

In our case study, we have made an explicit effort to compute and differentiate failures from faults. 
We have shown there is no correlation between failure efficiency and fault efficiency. 
Some of the prior works tend to consider either failures or faults, but very rarely both. There are actually very good reasons for that. 
On the one hand, the identification of failures requires to execute the configurations in real settings -- the process can be very costly and hard to engineer even for a scientific experiment. 
On the other hand, some works succeed to produce and collect many failures but ignore the actual correspondences with faults.  

Though we understand the underlying reasons and difficulties, our point is that the assessment of sampling techniques may highly vary depending on the metric considered (failure or fault efficiency). For example, all-one-enabled has a failure efficiency of 37.26\%  but is one of the worst techniques in terms of fault rate due of its high sample size. 
Our recommendation for researchers is to properly report and investigate the distinction (if any) between failures and faults. It is actually an open research direction to further characterize this distinction in other software engineering contexts than JHipster.





\subsubsection{Fault corpus} 


For the assessment of sampling techniques, one need is to define a correspondence between configurations and faults. As reported in the literature and in this study, this task is not trivial, time-consuming, and error-prone. 
A typical attitude is to manually build a corpus of faults with results confirmed by the developers, or from issues reported in mailing list or bug tracking systems.
For example, S\'anchez \etal performed  extensive analysis of Drupal's Git repository and issue tracking system to identify faults. 
A possible and important threat is that the corpus of faults is \emph{incomplete}. It can bias the empirical results since some faults may not be considered in the study. 

In our case study, we had a unique opportunity to collect \emph{all} faults through the testing of all configurations. Meanwhile we have been able to check whether these faults have been reported by JHipster developers. Our findings show that 6 important faults have been reported (see Table~\ref{tab:arules}). 
Though some of the faults were missing and required a manual investigation, they only impact a few configurations comparatively to faults reported on GitHub issues.

Overall, our findings suggest that a corpus of faults coming from an issue tracking system is a good approximation of the real corpus of faults. It is a positive result for other studies based on a manually collected corpus. 

\vspace{0.3cm}
\begin{tcolorbox}
\textbf{(RQ2.3)} \textit{How do our findings w.r.t. sampling effectiveness compare to other studies?}

\begin{inparaenum}[(i)]
\item \textbf{From a practical point of view}: We concur with previous findings that show that most-enabled-disabled is an interesting candidate to initiate the testing of configurations. For identifying further faults (and possibly all), we confirm that 2-wise or 3-wise provides a good balance between sampling size and fault-detection capability.
\item \textbf{From a researcher point of view}: Our results show that the assessment of sampling techniques may highly vary depending on the metrics used (failure or fault efficiency). Besides, a corpus of faults coming from an issue tracking system (GitHub) is a good approximation of the real, exhaustive corpus of faults. It is reassuring for research works based on a manually collected corpus. 
\end{inparaenum}
\end{tcolorbox}

%% file: parts_viewpoint.tex
\section{Practitioners Viewpoint (RQ3)}
\label{sec:practitioner}

We interviewed the JHipster lead developer, Ju\-lien Du\-bois, for one hour and a half, at the end of January. We prepared a set of questions and performed a semi-structured interview on Skype for allowing new ideas during the meeting. We then exchanged emails with two core developers of JHipster, Deepu K Sasidharan and Pascal Grimaud. Based on an early draft of our article, they clarified some points and freely reacted to some of our recommendations.   We wanted to get insights on how JHipster was developed, used, and tested.  We also aimed to confront our empirical results with their current practice.

\subsection{JHipster's Testing Strategy}
\label{subsec:jhipsterTesting}

\subsubsection{Continuous testing} 

JHipster relies on a continuous integration platform (Travis) integrated into GitHub. At the time of the release 3.6.1, the free installation of Travis allowed to perform 5 different builds in parallel, at each commit. JHipster exploits this feature to only test 12 configurations. 
JHipster developers give the following explanations: \emph{``The only limit was that you can only run 5 concurrent jobs so having more options would take more time to run the CI and hence affect our turn around hence we decided on a practical limit on the number [...] We only test the 12 combinations because we focus on most popular options and leave the less popular options out."}
Julien also mentioned that his company IPPON provides some machines used to perform additional tests. We can consider that the testing budget of JHipster 3.6.1 was limited to 12 configurations. It has a strong implication on our empirical results: \emph{Despite their effectiveness, some sampling strategies we have considered exceed the available testing budget of the project}. For example, a 2-wise sample has 41 configurations and is not adequate. A remaining solution is dissimilarity sampling (PLEDGE) of 12 configurations, capable of finding 5 failures and 3 faults.

\subsubsection{Sampling strategy} 

How have these 12 configurations been selected? According to Julien, it is both based on \emph{intimate technical knowledge of the technologies} and \emph{a statistical prioritization approach}. Specifically, when a given JHipster installation is configured, the user can send anonymous data to the the JHipster team so that it is possible to obtain a partial view on the configurations installed. 
The most popular features have been retained to choose the 12 configurations. For example, this may partly explain that configurations with Gradle are buggier than those with Maven -- we learned that Gradle is used in less than 20\% of the installations. There were also some discussions about improving the maintenance of Gradle, due to its popularity within a subset of contributors. The prioritization of popular configurations is perfectly understandable. Such a sample has the merit of ensuring that, at each commit, popular combinations of features are still valid (acting as non-regression tests). 
However, corner cases and some feature interactions are not covered, possibly leading to high percentage of failures. 

\vspace{0.3cm}
\begin{tcolorbox}
\textbf{(RQ3.1)} \textit{What is the most cost-effective sampling strategy for JHipster?}
Exhaustive testing sheds a new light on sampling techniques: 
\begin{inparaenum}[(i)]
\item the 12 configurations used by the JHipster team do not find any defect;
\item yet, 41 configurations are sufficient to cover the 5 most important faults;
\item dissimilarity and t-wise sampling  are the most effective.
\end{inparaenum}
\end{tcolorbox}

\subsection{Merits and Limits of Exhaustive Testing}

Julien welcomed the initiative and was seriously impressed by the unprecedented engineering effort and the 36\% failures. 
We asked whether the version 3.6.1 had special properties, perhaps explaining the 36\% of failures. He refuted this assumption and rather stated that the JHipster version was a major and stable release.  We explained that most of the defects we found were reported by the JHipster community.  The lead developer was aware of some interactions that caused problems in JHipster. These are known mostly from experience and not via the application of a systematic process. However, he ignored the significance of the failures. The high percentage of failures we found should be seriously considered since a significant number of users may be impacted given the popularity of the project. Even if faults involve rarely used configurations, he considered that the strength of JHipster is precisely to offer a diverse set of technologies. The effort of finding many failures and faults is therefore highly valuable. 

We then discussed the limits of testing all configurations. The cost of building a grid/cluster infrastructure is currently out of reach for the JHipster open-source project, due to the current lack of investments. JHipster developers stated: \emph{``even if we had limitless testing infrastructure, I do not think we will ever test out all possible options due to the time it would take"}. This observation is not in contradiction with our research method. Our goal was not to promote an exhaustive testing of JHipster but rather to investigate a cost-effective strategy based on collected evidence. 

Another important insight is that \emph{``the testing budget was more based on the time it would take and the resource it would use on a free infrastructure. If we let each continuous integration build to run for few hours, then we would have to wait that long to merge pull request and to make releases etc. So, it adds up lag affecting our ability to release quickly and add features and fixes quickly. So, turn around IMO is something you need to consider for continuous integration".} 

Finally, Julien mentioned an initiative\footnote{\url{https://github.com/jhipster/jhipster-devbox}} to build an all-inclusive environment capable of hosting any configuration. It is for JHipster developers and aims to ease the testing of a JHipster configuration on a local machine. In our case, we built a similar environment with the additional objective of automating the test of configurations. We have also validated this environment for all configurations in a distributed setting. 

\subsection{Discussions}

On the basis of multiple collected insights, we discuss trade-offs to consider when testing JHipster and address \textbf{RQ3}. 

\subsubsection{Sampling strategy} 

Our empirical results suggest using a dissimilarity sampling strategy in replacement to the current sampling based on statistical prioritization. It is one of the most effective strategy for finding failures and faults and it does not exceed the budget. In general, the focus should be on covering as much feature interactions as possible. If the testing budget can be sufficiently increased, t-wise strategies can be considered as well. However, developers remind us that \emph{``from a practical standpoint, a random sampling has possibility of us missing an issue in a very popular option thus causing huge impact, forcing us to make emergency releases etc., where as missing issues in a rarely used option does not have that implication"}. This applies to t-wise and dissimilarity techniques as well. Hence, one should find a trade-off between cost, popularity, and effectiveness of sampling techniques. We see this as an opportunity to further experiment with multi-objective techniques~\cite{Sayyad2013c,Parejo2016,Henard2015}. 

\subsubsection{Sampling size} 

Our empirical results and discussions with JHipster developers suggest that the testing budget was simply too low for JHipster 3.6.1, especially when popular configurations are included in the sampling.
According to JHipster developers, the testing budget \emph{``has increased to 19 now with JHipster 4, and we also have additional batch jobs running daily tripling the number of combinations [...] We settled on 19 configurations to keep build times within acceptable limits"\footnote{Discussions are available at \url{https://github.com/jhipster/generator-jhipster/issues/4301}}.}

An ambitious and long-term objective is to crowdsource the testing effort with contributors. Users can lend their machines for testing some JHipster configurations while a subset of developers could also be involved with the help of dedicated machines. 
In complement to continuous testing of some popular configurations, a parallel effort could be made to seek failures (if any) on a diversified set of configurations, possibly less popular. 

\subsubsection{Configuration-aware testing infrastructure} 

In any case, we recommend developing and \emph{maintain} a configuration-aware testing infrastructure. Without a ready-to-use environment, contributors will not be able to help in testing configurations. It is also pointless to increase the sample if there is no automated procedure capable of processing the constituted configurations. The major challenge will be to follow the evolution of JHipster and make the testing tractable. A formal model of the configurator should be extracted for logically reasoning and implementing random or t-wise sampling. New or modified features of JHipster should be handled in the testing workflow; they can also have an impact on the tools and packages needed to instrument the process.   

\vspace{0.3cm}
\begin{tcolorbox}
\textbf{(RQ3.2)} \textit{What are the recommendations for the JHipster project?}
To summarise, recommendations (and challenges) are:
\begin{inparaenum}[(i)]
\item for a budget of 19 configurations, dissimilarity is the most appropriate sampling strategy;
\item the trade-off between cost, popularity, and effectiveness suggests to further experiment with multi-objective techniques;
\item crowdsourcing the testing effort would help to face the computational cost of testing JHipster;
\item the development and maintenance of a configuration-aware testing infrastructure is mandatory to automate JHipster testing.
\end{inparaenum}
\end{tcolorbox}

%% file: parts_threats.tex
\section{Threats to Validity}

\label{sec:threats}

Our engineering effort has focused on a single but industrial and complex system. We expect more insights into characteristics of real-world systems than using diverse but smaller or synthetic benchmarks. With the possible consideration of all JHipster configurations, we gain a ground truth that allows us to precisely assess sampling techniques.

Threats to internal validity are mainly related to the quality of our testing infrastructure. 
An error in the feature model or in the configuration-aware testing workflow can typically produce wrong failures. We also used the Java and JavaScript tests generated by JHipster, as well as the data from only one entity model template (the one provided by the JHipster team). As reported, the validation of our solution has been a major concern during 8 man-months of development. We have used several strategies, from statistical computations to manual reviews of individual failures to mitigate this threat. Despite those limitations, we found all faults reported by the JHipster community and even new failures.

For the other remaining 242 configurations that fail due to our test infrastructure (see Section \ref{sec:infraissues}), there might be false positives. Since they only represent 0.9\% of all JHipster configurations, such false positives would have a marginal incidence on the results. In fact, this situation is likely to happen in a real (continuous integration) distributed testing environment (e.g., as reported in~\cite{Beller:2017:OMT:3104188.3104232}). We thus decided to keep those configurations in the dataset. Additionally, they can serve as a baseline to improve our testing infrastructure for the next versions of JHipster. 

To mitigate the threat related to missed studies comparing findings of configuration sampling techniques, we used a snowballing approach. We started from mapping studies and systematic literature reviews known by the authors. Selected studies have been reviewed by at least three authors.

%% file: parts_conclusion.tex
\section{Conclusion and Perspectives}
\label{sec:conclusion}

In this article, we reported on the first ever endeavour to test all configurations of an industrial-strength, open-source generator: JHipster. We described the lessons learned and assessed the cost of engineering a con\-fi\-gu\-ra\-tion-aw\-are testing infrastructure capable of processing 26,000+ configurations. 

\subsection{Synthesis of lessons learned}

\paragraph{Infrastructure costs.}
Building a configuration-aware testing infrastructure for JHipster requires a \textbf{substantial effort both in terms of human and computational resources}: \textbf{8 man-months} for building the infrastructure and \textbf{4,376 hours of CPU time} as well as \textbf{5.2 terabytes} of disk space used to build and run JHipster configurations. The most difficult part of realising the infrastructure was \textbf{to validate it}, especially in a distributed setting. These costs are system-dependent: for example, the Linux project provides tools to compile distinct random kernels, which can be used for various analyses (e.g., \cite{Melo:2016:QAV:2866614.2866615,DBLP:conf/icse/HenardPPKT04}), and ease the realisation of a testing infrastructure.       

\paragraph{Comparing sampling techniques.}
Almost \textbf{36\%} of the 26,0000 configurations fail. More than \textbf{99\%} of these failures can be attributed to \textbf{six interaction faults up to 4-wise}. The remaining failures are false positives.
As a result, in our case, t-wise testing techniques provide guarantees to \textbf{find all the faults}. Nevertheless, such guarantees come at a \textbf{price}, i.e., the number of configurations to sample (126). Still, only a \textbf{small subset} of the total number of configurations is necessary, validating the relevance of sampling techniques. Dissimilarity sampling is slightly better at finding failures though offering \textbf{generally good efficiencies w.r.t t-wise with a flexible budget}.  Most-enabled-disabled can be\textbf{very efficient regarding the very small number of configurations} it requires but should incorporate \textbf{randomness} in the sampling algorithm to not rely on a SAT solver's internal order \cite{Henard2014a}. Indeed, random sampling remains \textbf{a strong baseline} for failure and faults.  Finally, investigation of \textbf{both} faults and failures efficiencies shows that \textbf{they are not correlated} and that it is difficult to optimise them for a single sampling technique. Without the effort of testing all configurations, we would have missed important lessons or have superficially assessed existing techniques.   

\paragraph{Comparison with other studies.}
Our assessment of sampling techniques on JHipster confirm findings of the literature: \textbf{most-enabled-disabled is a relevant technique} in order to initiate testing, while t-wise techniques with low values of t provide interesting fault and failure detection ratios.

However, ranking sampling techniques \textbf{is highly sensitive to the metrics considered}, which \textbf{complicates} the ``\textit{comparison of comparing studies}". Yet, fault corpora issued from issue tracking systems such as GitHub \textbf{seem to contain almost all issues}, adding relevance to fault mining efforts.

\paragraph{Comparison with JHipster team testing practice.}
Confronting our results to the core JHipster developers was extremely valuable. First, the testing budget is \textbf{key}: even for 26,000+ configurations the developers did not have the possibility to test \textbf{more than twelve of them} continuously. Additionally, their strategy based on popular configurations \textbf{did not find any fault}. Such a stringent budget is a challenge for sampling techniques and combinatorial interaction testing ones in particular. In this case, dissimilarity is \textbf{our recommendation}. \textbf{Cost, popularity and fault-finding abilities} appeared as important factors in the determination of samples. This pushes for experimenting with mul\-ti-ob\-je\-cti\-ve techniques in such contexts. Finally, our effort in providing \textbf{an automated configuration-aware testing infrastructure is mandatory} for viable JHipster testing.       

\subsection{Test them all, is it worth it?}
Our investigations allow us to answer to the key question we raised in the title of this article: \textit{Is it worth testing all configurations?} Obviously, there is no universal `yes' or `no' answer to this question as it typically depends on the audience to which the question is asked. 
From a \textbf{researcher's perspective}, the answer is definitely `yes'. This effort enabled us to obtain a \textit{ground truth} notably on the faults for this specific version of JHipster. Sampling techniques can then be compared with respect to an absolute value (\textbf{\underline{all}} the faults), which is a stronger evidence than a comparison on a configuration subset. Building and running configurations also gave insights on failures that are not frequently analysed. This enthusiasm should be tempered with respect to the high costs of building a testing infrastructure capable of handling all configurations. Finally, JHipster has the interesting properties of being widely used as an open-source stack, non-trivial but still manageable in terms of configurations, which made this study possible. For example, researchers working on Linux kernels cannot envision to answer this question in the current state of computing since Linux has thousands of options~\cite{Nadi2015,Abal:2014,Melo:2016:QAV:2866614.2866615}.
From a \textbf{practitioner's perspective}, the answer is a resounding `no'. The JHipster community cannot afford computing and human costs involved with such an initiative. In the improbable event of having the sufficient resources, validation time at each release or commit would still be a no-go, point that is also likely to hold in other cases. Moreover, we have shown that sampling 126 configurations (out of 26,000+) is enough to find all the faults. While the absolute ranking between sampling methods is variable amongst studies and cases analysed, sampling techniques are more efficient at faults and failures than exhaustive testing, as illustrated by the poor 0.02\% of fault efficiency when sampling all configurations. 
Though testing all configurations is not worth it, we recommend to develop a testing infrastructure \emph{able} to handle all possible configurations; it is a mandatory prerequisite before instrumenting a cost-effective sampling strategy. 


\subsection{Perspectives}

Our empirical study opens opportunities for future work, both for practitioners and researchers. Future work will cover fitting the test budget in continuous integration setting and devise new statistical selection/prioritisation sampling techniques, including how techniques like association rules work in such settings. We plan to continue our collaboration with the JHipster community. Contributors involvement, testing infrastructure evolution and data science challenges (e.g., \cite{Kim2016}) are on the agenda. 
Our long-term objective is to provide evidenced-based theories and tools for continuously testing configurable systems.
